\documentclass[fleqn, usenatbib]{mnras}

\usepackage[colorinlistoftodos]{todonotes}
\usepackage{graphicx}	
\usepackage{amsmath}	
\usepackage{amssymb}	
\usepackage{aas_macros}
\usepackage{longtable}
\usepackage{color}
\usepackage{dcolumn}	
\usepackage{multirow}
\usepackage[english]{babel}
\usepackage{color}
\usepackage{hyperref}

\definecolor{ao(english)}{rgb}{0.0, 0.5, 0.0}

\newcommand{\msun}{${\rm M}_{\odot}\;$}

\newcommand*{\ditto}{\texttt{"}}

\title[Searches for GC binary pulsars and fast transients]{Multi-epoch searches for relativistic binary pulsars and fast transients in the Galactic Centre}

\author[R.~P.~Eatough et al.]{\parbox{\textwidth}{R.~P.~Eatough,$^{1,2}$\thanks{E-mail: reatough@nao.cas.cn}
P. Torne,$^{3,2}$\thanks{E-mail: torne@iram.es} 
G.~Desvignes,$^{4,2}$
M.~Kramer,$^{2,5}$
R.~Karuppusamy,$^{2}$
B.~Klein,$^{2,6}$
L.~G.~Spitler,$^{2}$
K.~J.~Lee,$^{7}$
D.~J.~Champion,$^{2}$
K.~Liu,$^{2}$
R.~S.~Wharton,$^{2}$
L.~Rezzolla$^{8,9,10}$
and H.~Falcke$^{11,2}$
}
\vspace{0.4cm} \\ 
\parbox{\textwidth}{
$^{1}$National Astronomical Observatories, Chinese Academy of Sciences, 20A Datun Road, Chaoyang District, Beijing 100101, P.~R.~China\\
$^{2}$Max-Planck-Institut f\"{u}r Radioastronomie, Auf dem H\"{u}gel 69, D-53121, Bonn, Germany\\
$^{3}$Institut de Radioastronomie Millim\'etrique (IRAM), Avda. Divina Pastora 7, N\'{u}cleo Central, 18012, Granada, Spain\\
$^{4}$Laboratoire d'\'{E}tudes Spatiales et d'Instrumentation en Astrophysique, Observatoire de Paris, Universit\'{e} Paris-Sciences-et-Lettres, Centre National de la Recherche Scientifique, Sorbonne Universit\'{e}, Universit\'{e} de Paris, 5 place Jules Janssen, 92195 Meudon, France\\
$^{5}$Jodrell Bank Centre for Astrophysics, School of Physics and Astronomy, The University of Manchester, Manchester M13 9PL, UK\\
$^{6}$University of Applied Sciences Bonn-Rhein-Sieg, Sankt Augustin, Germany\\
$^{7}$Kavli Institute for Astronomy and Astrophysics, Peking University, Beijing 100871, P.~R.~China\\
$^{8}$Institut f\"{u}r Theoretische Physik, Goethe-Universit\"{a}t, Max-von-Laue-Straße 1, D-60438 Frankfurt, Germany\\
$^{9}$Frankfurt Institute for Advanced Studies, Ruth-Moufang-Straße 1, 60438 Frankfurt, Germany\\
$^{10}$School of Mathematics, Trinity College, Dublin 2, Ireland\\
$^{11}$Department of Astrophysics, Institute for Mathematics, Astrophysics and Particle Physics (IMAPP), Radboud University, P.O. Box 9010, 6500 GL Nijmegen, The Netherlands
}
}

\date{Accepted XXX. Received YYY; in original form ZZZ}

\begin{document}
\label{firstpage}
\pagerange{\pageref{firstpage}--\pageref{lastpage}}
\maketitle
\begin{abstract}
The high stellar density in the central parsecs around the Galactic
Centre makes it a seemingly favourable environment for finding
relativistic binary pulsars. These include pulsars orbiting other
neutron stars, stellar-mass black holes or the central supermassive
black hole, \mbox{Sagittarius$\,$A*}. Here we present multi-epoch
pulsar searches of the Galactic Centre at four observing
frequencies, $(4.85,\,8.35,\,14.6,\,18.95)\,{\rm GHz}$, using the \mbox{Effelsberg$\,$100-m} radio telescope. 
Observations were conducted one year prior to the discovery of, and during monitoring observations of, the Galactic Centre magnetar PSR~J1745$-$2900. Our data analysis features acceleration searches on progressively shorter time series to maintain sensitivity
to relativistic binary pulsars. The multi-epoch observations increase the likelihood of discovering transient or nulling pulsars, or ensure
orbital phases are observed at which acceleration search methods work optimally. In $\sim147\,{\rm h}$ of separate observations,
no previously undiscovered pulsars have been detected. Through
calibration observations, we conclude this might be due to insufficient instantaneous
sensitivity; caused by the intense continuum emission from the Galactic Centre, its large distance and, at higher
frequencies, the aggregate effect of steep pulsar spectral indices and
atmospheric contributions to the system temperature. Additionally we find that for millisecond pulsars in wide circular orbits $(\lesssim800\,{\rm d})$ around \mbox{Sagittarius$\,$A*}, linear acceleration effects
cannot be corrected in deep observations $(9\,{\rm h})$ with existing software tools. 
Pulsar searches of the Galactic Centre with the next generation of
radio telescopes $-$ such as MeerKat, ngVLA and SKA1-mid $-$ will have improved chances of uncovering this elusive population.
\end{abstract}



\section{Introduction}
Binary radio pulsars are precision tools for tests of gravitational
theories in the strong field regime \citep[see e.g.][for a review of
  key results]{wex+2014}. In general, the larger the mass of the
pulsar companion, and the more compact and eccentric its orbit, 
increased is the extent and precision to which gravity tests can be
performed. For this reason, the Galactic Centre (GC) is a tantalizing
target for pulsar searches. In addition to having the highest stellar
density in the Galaxy, it hosts the massive compact object,
Sagittarius~A* (hereafter Sgr~A*), which is shown to be a
supermassive black hole (SMBH). With a mass of approximately
$4\times10^{6}\,$\msun and a distance just over $\rm{8\,kpc}$
\citep{eck96,gsw+08,gill09,grvty+III} it is Earth's nearest SMBH and offers a
unique opportunity to test the General Theory of Relativity (GR) and
the base properties of black holes via astrometry of orbiting stars
\citep{grvty+I,Doeaav8137} and direct VLBI imaging at mm-wavelengths
\citep{eht+I}. The detection of even a ``typical'' pulsar  (spin period, $P\sim0.5\,{\rm s}$) in an orbit
around Sgr~A*, of the order of years, can enable tests of the Cosmic
Censorship Conjecture and the No Hair Theorem; two of the most
fundamental predictions of GR \citep{wk99,kra04,liu12,le+17}. When combined with
stellar astrometry and imaging, results from a pulsar experiment are
highly complementary and will help to build a complete description of
Sgr~A* \citep{pwk+16}.

Multi-wavelength observations of the GC indicate that the number
of pulsars in the central few parsecs should be high
\citep{whar12} and conditions are highly favourable for relativistic binaries \citep{fgl+11}. The dense nuclear star cluster surrounding Sgr~A*
\citep[see e.g.][for a review]{gen10} contains a majority of older
late-type stars, but contrary to expectations, massive
young main-sequence stars \citep{gdm+03} and possible neutron star progenitors such as
Wolf-Rayet stars \citep{paumard+01}. The presence of neutron stars is
further indicated by large numbers of X-ray binaries, possible pulsar
wind nebulae, X-ray features like the ``cannonball'' and compact radio variables 
\citep{muno+05,wang+06,zhao+13,zmm+20}. Despite this only
six radio pulsars have been discovered within half a degree of Sgr~A*
\citep{john06,den09,efk+13,sj13} even after many dedicated searches at multiple 
wavelengths \citep{kjm+96,kkl+00,kkm+04,Klein:thesis,Deneva:thesis,mq10,sbb+13,ekk+13}. 
Hyperstrong scattering of radio waves in the GC has been  
the principal explanation for the scarcity of detected pulsars \citep{cl97,lc1+98,lc2+98,cl+02}, however, 
scatter broadening measurements of PSR~J1745$-$2900 in \cite{spi14} and \cite{bow14} appear to 
contest this\footnote{\cite{spi14} note that if
  the scatter broadening observed in PSR~J1745$-$2900 is
  representative of the GC as a whole, millisecond pulsars (MSPs)
  would still remain undetectable at low frequencies viz. at frequencies
  $\lesssim 7\,{\rm GHz}$. Indeed \cite{macq15} argue that the GC
  pulsar population is likely dominated by MSPs that still necessitate such high
  frequency searches.}. Other authors have noted that the lack of GC pulsars is expected
  under a certain set of conditions and considering the sensitivity limits of existing pulsar surveys
\citep{cl+14,rla+17,le+17}. Alternatively, the scarcity of detected pulsars might be caused by a 
more complex scattering structure toward the GC \citep{cordes1997,lc1+98,lc2+98,john06,sef+16,ddb+17}.

In this work we have tackled additional and necessary
requirements for pulsar searches of the GC that have thus far not been fully
addressed. These include: a) repeated high frequency $(\gtrsim5\,{\rm
  GHz})$ observations over a long duration $(\mathcal{O}\,{\rm yr})$; b) analysis of all observations with binary pulsar search
algorithms capable of detecting pulsars in binary systems with a wide range
of orbital periods; c) searches for bright single pulse emission from fast transient sources. 
The following is a brief outline of the rest of this paper.  In
Section~2 the observations, data processing and measurements of the
observational system sensitivity are described. In Section~3 the basic
results of this search are given. Section~4 provides discussion of our results in the context of
previous pulsar searches of the GC and remaining shortcomings. We also discuss the prospects
for future searches of the GC with current and next generation radio telescopes. Section~6 presents a summary.

\begin{figure}
    \begin{center}
        \includegraphics[width=\columnwidth]{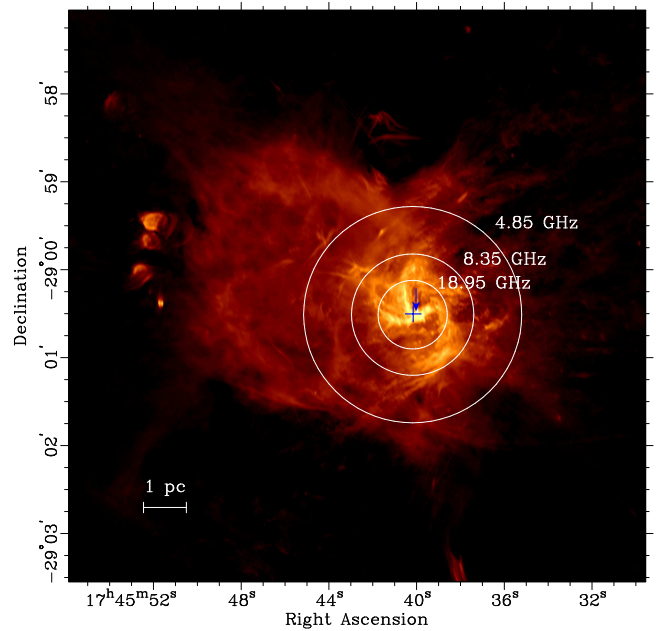}
        \caption[]{The areas covered by our search at three observing frequencies as indicated by the relevant HPBW overlaid on a 5.5$\,$GHz Jansky VLA map of the \mbox{Sgr$\,$A} complex from \citet{zmg16}. The HPBW at 14.6$\,$GHz is only 10 per cent larger than that at 18.95$\,$GHz, so has not been plotted for clarity. The blue cross marks the position of PSR~J1745$-$2900 and the location of \mbox{Sgr$\,$A*} is indicated by the blue vertical arrow in the top right quadrant of the cross. A physical scale of 1~pc is indicated in the bottom left using a recently derived geometric GC distance \citep{grvty+III}.}
    \label{f:region}
    \end{center}
\end{figure}

\section{Observations, Data Processing and System Sensitivity}\label{sec:obs_analysis}
\subsection{Observations}
Observations were made with the Effelsberg$\,$100-m radio telescope,
of the Max Planck Institute for Radio Astronomy, between February~2012
and February~2016. In 2012 all data were taken at 18.95$\,$GHz and
targeted the position of \mbox{Sgr$\,$A*} ($\rm{RA_{J2000} = 17^{\rm
    h}\,45^{\rm m}\,40^{\rm s}.0409}$, $\rm{Dec_{J2000} =
  -29^{\circ}\,00\arcmin\,28\arcsec.118}$; \citealt{rb+2004}), as
part of a `stack search' for solitary pulsars \citep{ekk+13}. After
the discovery of radio pulsations from the GC magnetar
PSR~J1745$-$2900 in April 2013 \citep{efk+13, sj13}, a multi-frequency
monitoring program of this object was started (see \citealp[][for
  details of this campaign]{dep+18}). Pulsar search-mode observations
were performed in parallel with timing 
observations of PSR~J1745$-$2900 that were folded with a contemporaneous
ephemeris. All observations were centred on the \mbox{X-ray} position
of PSR~J1745$-$2900 as measured with the Chandra observatory
\citep[$\rm{RA_{J2000} = 17^{\rm h}\,45^{\rm m}\,40^{\rm s}.2}$,
  $\rm{Dec_{J2000} = -29^{\circ}\,00\arcmin\,30\arcsec.4;}$
][]{rea13}, 2.4$\arcsec$ away from \mbox{Sgr$\,$A*}. Receivers with central frequencies, $\nu$, of
$(4.85,\,8.35,\,14.60,\,18.95)\,{\rm GHz}$, and respective half-power beam widths (HPBW) of
146$\arcsec$~(5.8$\,$pc), 82$\arcsec$~(3.3$\,$pc),
51$\arcsec$~(2.0$\,$pc) and 46$\arcsec$~(1.8$\,$pc) $-$ where the
corresponding physical scale at the distance of the GC is indicated in
brackets $-$ were used (see Figure~\ref{f:region}). At all frequencies
both PSR~J1745$-$2900 and \mbox{Sgr$\,$A*} are within the HPBW and any
reduction in sensitivity toward \mbox{Sgr$\,$A*} due to offset
pointing was negligible.

The search-mode data $-$ a digital filterbank with bandwidth 500$\,$MHz,
128 spectral channels and a sampling interval of 65.536\,$\mu$s $-$
were recorded with the Pulsar Fast-Fourier-Transform Spectrometer
(PFFTS) backend at the three lowest frequencies. At 18.95\,GHz the
X-Fast-Fourier-Transform Spectrometer (XFFTS) was used to cover the
larger available bandwidth of 2$\,$GHz. Here each linear polarization
was sampled independently with an interval of 128\,$\mu$s across 256 spectral
channels. After acquisition, the two polarisations were combined
offline. Data from both backends originally consisted of 32-bit
integer samples for each frequency channel in a bespoke data
format. This was converted to {\sc
  sigproc}\footnote{\url{http://sigproc.sourceforge.net}} {\sc filterbank}
format along with downconversion to 8-bit samples performed by a dedicated {\sc C++}
program.

In this work, 112 independent epochs (or $\sim$147$\,$h) of GC
observations have been taken and analyzed. The maximum duration of an
individual observation was 2.4$\,$h, which is the total time the Sgr~A* region is visible from Effelsberg each day. The benefits of
our repetitive observational scheme is manifold. Binary pulsars can be
``hidden'' by transient phenomena such as relativistic spin-precession
\citep[e.g.][]{k+98,bkk+08,dkl+19}, binary eclipses \citep[e.g.][]{jml92,f+05} and
sub-optimal orbital phases for acceleration search algorithms
\citep{ekl+13}. In addition, both solitary and binary pulsars can
exhibit burst-like or transient emission in the form of giant pulses
\citep[e.g.][]{jsk+05}, nulling or intermittent pulsations
\citep{klo+06,kek+13}. These effects can make it impossible to detect certain pulsars in just a single survey observation.

\subsection{Data processing} 
\label{GCsurvey:sec_data_analysis}
The data were processed using the Max-Planck-Gesellschaft
Supercomputer
HYDRA\footnote{\url{http://www.mpcdf.mpg.de/services/computing/hydra}}
with a pulsar searching pipeline utilizing the {\sc
  presto}\footnote{\url{https://www.cv.nrao.edu/~sransom/presto}} software
package \citep{ransom2001}. A basic outline of the search pipeline is
as follows.

Firstly, because pulse broadening caused by scattering toward the GC
\citep{spi14} made the original sampling interval unnecessarily fine,
data at 4.85\,GHz and 8.35\,GHz were down-sampled in time, with a
dedicated {\sc python} script, by a factor of four and two
respectively; thereby reducing the computational requirements. At the
higher frequencies of $14.6\,{\rm GHz}$ and $18.95\,{\rm GHz}$, where scatter
broadening is smaller, no down-sampling was performed.

\begin{table}
  \centering
    \caption{Details of the data configuration and searching parameters
    used at each central observing frequency, $\nu$. 
    $\rm{\Delta \nu}$ is the total bandwidth, $n_{\rm c}$ is the
    number of frequency channels, $T_{\rm obs}$ is the total
    integration length, $\tau$ is the sampling time, $\Delta {\rm DM}$
    is the range of dispersion measures explored, $n_{\rm seg}$ is the
    number of consecutive segments of the original integration
    analyzed in the acceleration search, viz. $2^0$ corresponds to one
    segment of the full integration length $T_{\rm obs}$, $2^1$ corresponds to two
    consecutive segments of length $T_{\rm obs}/2$ and so forth. At
    all frequencies and in all segments the $z_{\rm max}$ parameter, which
    denotes the maximum number of Fourier bin drifts searched in the {\sc
      presto} routine {\sc accelsearch}, was set to $z_{\rm max}=1200$.}
  \label{t:obsparams}
  \begin{tabular}{rcccrccc}
    \hline  
    \multicolumn{1}{c}{$\nu$}		& $\rm{\Delta \nu}$	& $n_{\rm c}$ & $T_{\rm obs}$ & \multicolumn{1}{c}{$\tau$}	
    & $\Delta{\rm DM}$ & \multicolumn{1}{c}{$n_{\rm seg}$}  \\
    \multicolumn{1}{c}{(GHz)}			& (GHz) 		&& (hr) & \multicolumn{1}{c}{($\rm{\mu s}$)} 	 
    & ($\rm{cm^{-3}\,pc}$) &	 	\\
    \hline
    \hline
    $\rm{4.85}$ 			& 0.5			&128 &1.2 & 262.1 & $800$-$11,900$      & $2^0$-$2^4$ 	\\
    $\rm{8.35}$ 			& 0.5			&128 &2.4 & 131.1 & $800$-$15,080$	& $2^0$-$2^5$ 	\\
    $\rm{14.60}$ 		& 0.5			&128 &1.2 &  65.5 & $800$-$10,800$	& $2^0$-$2^4$ 	\\
    $\rm{18.95}$			& 2.0			&256 &2.4 & 128.0 & $800$-$21,200$      & $2^0$-$2^5$  \\
    \hline
  \end{tabular}
\end{table}

Next the data from an individual observation (typically of length
$\rm{\sim1.2\,h}$ or $\rm{\sim2.4\,h}$) was recursively
split into segments of half the observing duration, $T_{\rm obs}$, down to
a minimum segment length of $\simeq4.5\,{\rm min}$. Segmentation of
the data is performed in order to maintain sensitivity to pulsars in
shorter orbital period systems using so-called ``constant acceleration
searches'' that are most effective when $T_{\rm obs}\lesssim P_{\rm
  b}/10$, where $P_{\rm b}$ is the orbital period
\citep{ransom03,ng15}. Such segmented acceleration search
schemes have been employed in a re-analysis of the Parkes multi-beam
pulsar survey \citep{ekl+13}, and to the low Galactic latitude region
of the High-Time-Resolution-Universe South survey \citep{ng15},
discovering the most highly accelerated pulsar currently known
\citep{cck+18}. From the relation described above, our minimum segment length of $\simeq4.5\,{\rm min}$ 
results in a minimum detectable orbital period of $\simeq45\,{\rm min}$.
However, segmentation of data to improve binary
search sensitivity is a trade-off with the intrinsic sensitivity,
defined by the minimum detectable flux density, $S_{\rm min}$, which
is $\propto\,\sqrt{T_{\rm obs}}$ (see Section~\ref{ss:sensitivity} for
more details of the sensitivity of the observing system used here). Hereafter $T_{\rm obs}$ can refer to the duration of the original observation, or the length of an individual data segment. 

After segmentation, the detrimental effects of Radio Frequency
Interference (RFI) were mitigated with the {\sc presto} routine {\sc
  rfifind} and a list of periodic signals to be excluded from further
analysis. This list includes the domestic mains power, at frequency
50$\,$Hz with a number of its harmonics, and the first 32 integer
harmonics of PSR~J1745$-$2900 which is prevalent throughout the
observations after April 28$^{\rm th}$ 2013.  The data were then
transformed into the inertial reference frame of the solar system
barycentre and dedispersed to time series assuming different
dispersive delays, starting at a trial dispersion measure (DM) of
800$\rm \,cm^{-3}\,pc$ based on the DM of known pulsars in the GC\footnote{Using the free electron density model in \citet{ymw+17}, this DM corresponds to a minimum distance of $5\,{\rm kpc}$.}, with the largest trial DM value and the number
and size of the DM steps dependent on the frequency, bandwidth and
sampling interval of the observation as determined by the {\sc presto} routine {\sc ddplan.py}\footnote{${\rm DM}=10,000\rm \,cm^{-3}\,pc$ was our chosen upper limit at all frequencies. The variable upper limits presented in Table~1 were caused by a scripting error but added little to the overall processing time so remained unchanged.}. For a DM of 800$\rm \,cm^{-3}\,pc$ (equivalent to our lowest choice of trial DM) 
intra-channel dispersion smearing at the lower frequency edge of the band is $266$ and $49\,\mu{\rm s}$ 
at $4.85$ and $8.35\,{\rm GHz}$ respectively. This corresponds to $1.01\tau$ and $0.19\tau$, where $\tau$ is the sampling time, at $4.85$ and $8.35\,{\rm GHz}$ respectively. Neglecting pulse scattering (which could be the dominant effect in this direction), 
a marginal reduction in sensitivity to MSPs might therefore be expected at $8.35\,{\rm GHz}$.
Details of the various observational 
configurations and data processing parameters used are summarised in Table~\ref{t:obsparams}.

A search for single pulses from fast transient sources was performed on the longest available dedispersed
time series using {\sc single\_pulse\_search.py} from {\sc presto}. Box-car filters with widths up to $150\tau$ 
were used to record events with intensity $\geq 6\,\sigma$ (see \citealp[e.g.][]{kakl+15} for details on single pulse searches).
The dedispersed time series were then corrected for red noise effects
and searched for periodic (and accelerated) signals with the {\sc
  presto} program {\sc accelsearch}. The line-of-sight (l.o.s) acceleration caused by a binary companion, $a$,
makes the spin frequency, $f$, of a pulsar drift in the Fourier
spectrum by a number of spectral bins, $n_{\rm drift}$, given by
$n_{\rm drift} = af{T_{\rm obs}}^2/c$, where $T_{\rm obs}$ is the integration length and $c$ is the speed of light. 
{\sc accelsearch} uses the `correlation technique'
to collect the smeared signal back into a single spectral bin by
the application of Fourier domain matched filtering; in practice by
convolution of a small range of Fourier bins, around the relevant
spectral bin, with the complex conjugated and frequency reversed
version of the finite impulse response (FIR) filter that describes the
signal smearing \citep{ransom02,dat+18}. The acceleration search,
which dominated the data processing time, was performed with the
Graphical Processing Unit (GPU) enabled version of the {\sc
  accelsearch}
routine\footnote{\url{https://github.com/jintaoluo/presto2\_on\_gpu}}. For
all segment lengths the maximum value of $n_{\rm drift}$ searched (given by
the {\sc accelsearch} input parameter $z_{\rm max}$) was $z_{\rm max}=1200$.
At this stage harmonic summing was also applied, with
up to 16 harmonics summed for non-accelerated signals, and up to 8
harmonics for highly accelerated signals. 

Results from the periodicity and acceleration searches were consolidated with the {\sc presto} 
routine, {\sc accel\_sift.py}, to leave only detection with a harmonically summed power 
$\geq 6\,\sigma$; removing in the process duplicated (i.e. detected at different DMs and
accelerations) and harmonically related signals. Given our choice of threshold (bespoke to these data),
typically $\gtrsim 100$ candidates per segment were then folded with {\sc presto prepfold} to create candidate 
evaluation plots. Lastly, visual inspection of these candidates was done manually via interactive scatter 
diagrams and/or automatically utilizing PICS AI and PEACE \citep{zhu14,lee13}.

\subsection{System Sensitivity}
\label{ss:sensitivity}

\begin{table*}
  \centering
  \caption{System sensitivity measurements $-$ indicated by $T_{\rm sys} -$ at all observing frequencies (first column) and on three sky positions described in Section~\ref{ss:sensitivity} (third, fourth and fifth columns). The various contributions to $T_{\rm sys}$ from the GC, $T_{\rm GC}$, elevation dependent ground spillover and atmospheric effects, $T_{\rm elv}$, and the receiver, $T_{\rm rec}$, are indicated in the sixth, seventh and eighth columns respectively. Errors given in brackets are derived from the standard deviation of measurements over one year at $4.85\,{\rm GHz}$ and $8.35\,{\rm GHz}$, and from the standard deviation of measurements of $T_{\rm sys}$ in the first 60 individual frequency channels at $14.6\,{\rm GHz}$. The origin of figures given at $18.95\,{\rm GHz}$, where calibration was not done, are described in Section~\ref{ss:sensitivity}. Values of the telescope gain, receiver temperature measured at the zenith angle, $T^{\dagger}_{\rm{rec}}$, and the $\sigma_{\rm min}=10$ minimum detectable flux density (assuming a pulse width of $0.05P$ and maximum $T_{\rm obs}$ as given in Table~\ref{t:obsparams}) in the Sgr~A* ON position, $S_{\rm min}$, are given in the second, ninth and tenth columns respectively.}
  \label{table:tsys_sefds_slims}
  \begin{tabular}{rccccrcccc}
  \hline
  \multicolumn{1}{c}{}	& \multicolumn{1}{c}{} & \multicolumn{1}{c}{Sgr A* ON} & \multicolumn{1}{c}{Sgr A* OFF} & \multicolumn{1}{c}{NGC7027 OFF} & \multicolumn{1}{c}{}   & & & & \multicolumn{1}{c}{Sgr A* ON}	\\
  
  \multicolumn{1}{c}{$\nu\,{\rm (GHz)}$} & \multicolumn{1}{c}{$G\,({\rm K \,Jy^{-1}})$} & \multicolumn{1}{c}{$T_{\rm{sys}}\,{\rm (K)}$} & \multicolumn{1}{c}{$T_{\rm{sys}}$ (K)} & \multicolumn{1}{c}{$T_{\rm{sys}}$ (K)} & \multicolumn{1}{c}{{$T_{\rm{GC}}$} (K)}	& \multicolumn{1}{c}{{$T_{\rm{elv}}$}(K)} & \multicolumn{1}{c}{{$T_{\rm{rec}}$}(K)} & $T^{\dagger}_{\rm{rec}}$(K) & \multicolumn{1}{c}{$S_{\rm min}$(mJy)}\\
 \hline
 \hline
   4.85    	& 1.55 & 200(16)	&  63(3)	&  31(3) & 137(16) & 32(4) &  31(3) & 27 & $0.14$			  	\\
   8.35    	& 1.35 & 126(23)	&  72(15)	&  31(8) &  55(27) & 41(17) &  31(8) & 22 & $0.07$			 	\\
  14.60    	& 1.14 & 194(16)	    & 155(13)	& 149(12) &  39(21)	&  6(18) & 149(12) & 99 & $0.19$			    \\
  18.95    	& 1.03 &  \multicolumn{1}{c}{$\sim$ 124}	& \multicolumn{1}{c}{$-$} & {$\sim$ 104}  &  \multicolumn{1}{c}{$20$} &  \multicolumn{1}{c}{$-$}	& \multicolumn{1}{c}{$\sim$ 104} & 64 & {0.05}                    \\    	
 \hline
 \end{tabular}
\end{table*}

\begin{figure}
    \begin{center}
        \includegraphics[height=\columnwidth,angle=-90]{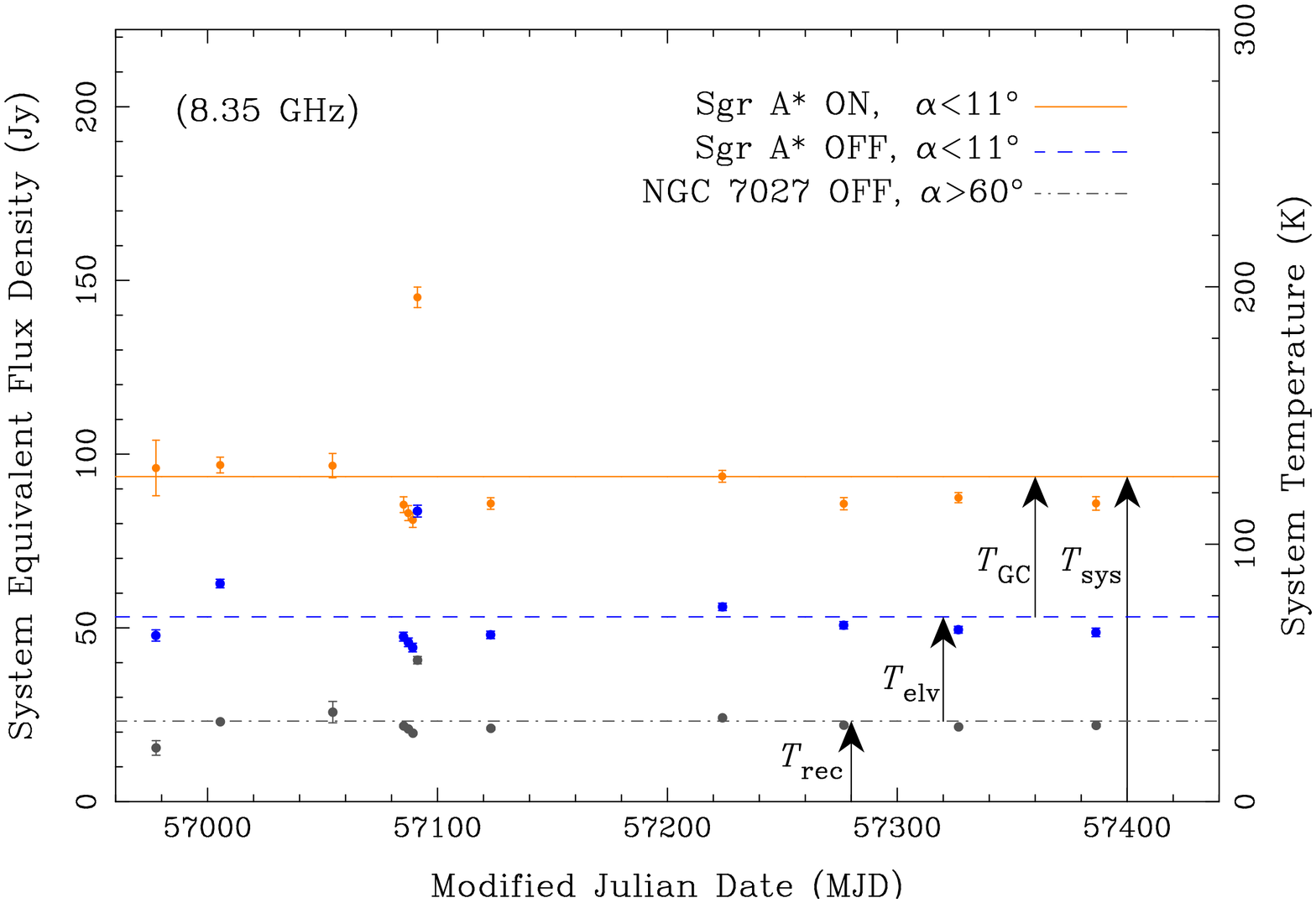} 
        \includegraphics[height=\columnwidth,angle=-90]{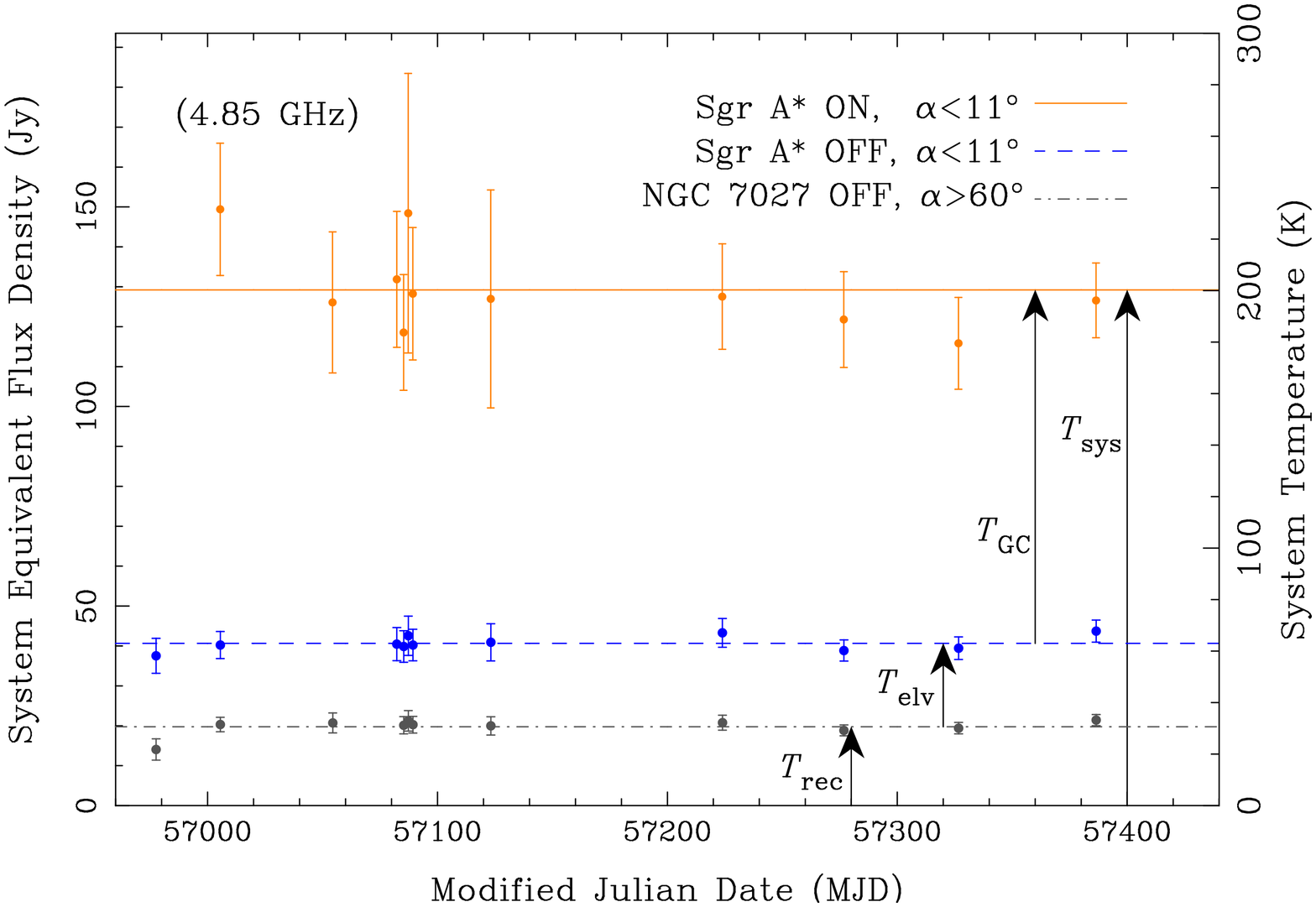}
        \caption[]{Observing system sensitivity $-$ quantified by the system
          temperature $T_{\rm sys}$, or the system equivalent flux
          density $S_{\rm sys}=T_{\rm sys}/G\;-$ over the course of
          one year of observations at Effelsberg. The top panel shows
          values measured at $8.35\,{\rm GHz}$ while the bottom panel
          shows those at $4.85\,{\rm GHz}$. Orange points indicate
          measurements toward Sgr~A$^*$; blue points are values
          measured at an equivalent telescope elevation as Sgr~A$*$
          ($\alpha<11^{\circ}$) but `off-source' and black points show
          measurements at high telescope elevation ($\alpha>60^{\circ}$). Error
          bars are given from the standard deviation of values across
          all channels in the observing band, and solid or dashed horizontal
          lines show the average values.}
    \label{fig:cal_sens}
    \end{center}
\end{figure}

To calculate flux density or luminosity limits of the searches
presented in this work, and to establish the major contributing factors
to reductions in the sensitivity of our observing system, we have
calibrated the sensitivity at the three lowest
frequencies. In all cases the planetary nebula ${\rm NGC}\,7027$, with a known
radio spectrum \citep{zavh+08}, was used as a reference source
and calibration was performed with the {\sc psrchive} software
package\footnote{\url{http://psrchive.sourceforge.net}}. At {18.95\,{\rm GHz}} no calibration 
observations were performed and the sensitivity was estimated from published system parameters.

From the radiometer equation, as applied to observations of pulsars,
the limiting flux density, $S_{\rm min}\,({\rm mJy})$, of a pulsar search
observation can be written
\begin{equation}
\label{eq:radiometer}
S_{\mathrm{min}} = \beta \frac{ {\sigma_{\rm min}} \, T_{\rm sys}}{G\sqrt{n_{\mathrm{p}}\,T_{\mathrm{obs}}\,B}}\sqrt{\frac{W}{P-W}},
\end{equation}
where $\beta$ accounts for digitisation losses and is negligible
$(\beta\simeq1)$ for 8-bit sampling, $G\,({\rm K\,Jy}^{-1})$ is the telescope gain,
${\sigma_{\rm min}}$ is the minimum statistically detectable signal to
noise ratio, $n_{\rm p}$ is the number of polarisations summed, $B\,({\rm MHz})$ is the receiver bandwidth
and $P$ and $W$ are the pulse period and width respectively. $T_{\rm
  sys}\,({\rm K})$ is the system temperature given by 
\begin{equation}
T_{\rm sys}=T_{\rm rec}+T_{\rm sky}+T_{\rm atm}+T_{\rm elv}. 
\end{equation}
Here $T_{\rm rec}$ is the
instrumental receiver noise temperature, $T_{\rm sky}$ is the
astrophysical background sky temperature, $T_{\rm atm}$ is the
combined effects of atmospheric opacity and water vapour emission (centred around $\sim22\,{\rm GHz}$) and $T_{\rm elv}$ is the
elevation dependent blackbody spillover radiation from the ground combined with an increased column depth of atmosphere. As
noted by \citet{john06} and \citet{mq10}, at frequencies $\lesssim8\,{\rm GHz}$, $T_{\rm sky}$ has a significant impact on the sensitivity of GC pulsar
searches because continuum emission in the GC $-$ from a combination
of thermal and nonthermal sources $-$ is known to be exceptionally
bright: $T_{\rm sky} \sim T_{\rm GC}\simeq\mathcal{O}100\,{\rm K}$
\citep[e.g.][]{pae+89,rfr+90,lzc+08}. At frequencies $\gtrsim15\,{\rm
  GHz}$, atmospheric and elevation dependent spillover effects are
thought to become the dominant contributors to $T_{\rm sys}$ \citep{mq10}.

\begin{figure}
    \begin{center}
    \includegraphics[scale=0.35,angle=-90]{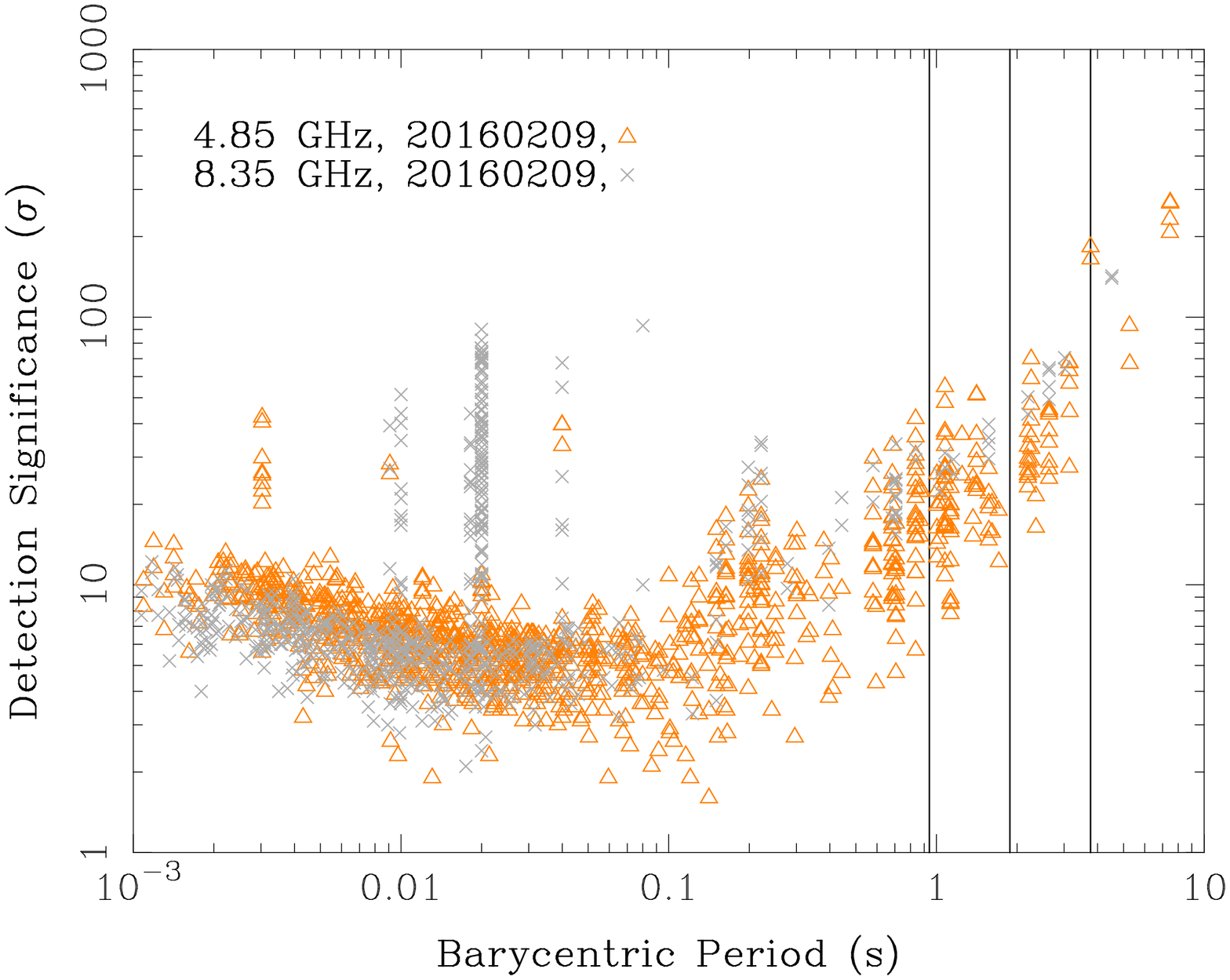}
    \includegraphics[scale=0.35,angle=-90]{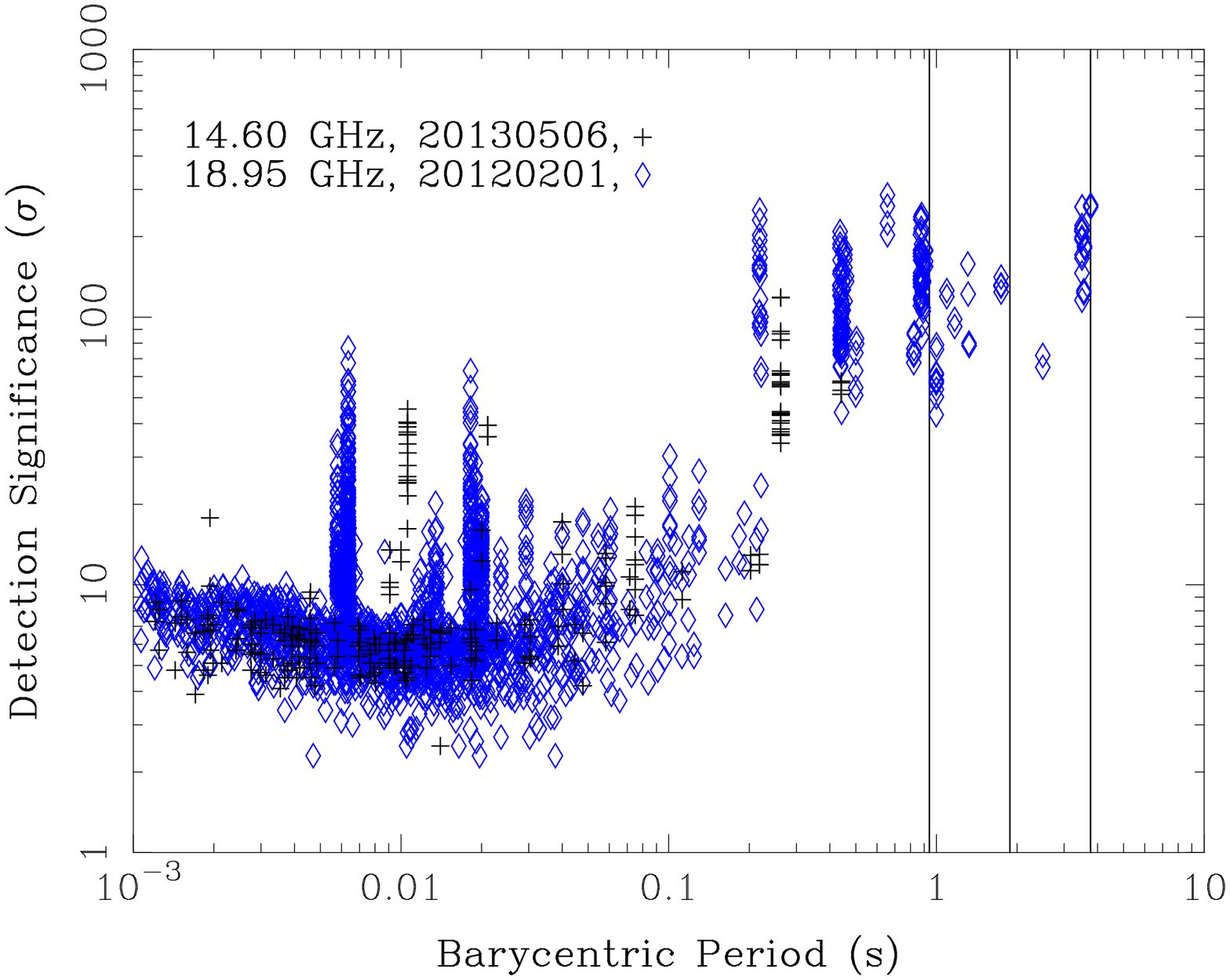}
    \caption{Scatter diagrams showing the barycentric pulse period versus folded detection significance of pulsar candidates generated from the periodicity search of four individual epochs (dates in legend) at different observing frequencies. Each point represents a pulsar candidate that was inspected either by eye and/or with machine learning tools. The upper panel shows results from searches at $4.85\,{\rm GHz}$ and $8.35\,{\rm GHz}$ and the lower panel those at $14.6\,{\rm GHz}$ and $18.95\,{\rm GHz}$. The solid vertical lines indicate the barycentric spin period of PSR~J1745$-$2900 and subsequent harmonics at one half and one quarter of the fundamental spin period.} 
    \label{Fig:scatter_diag}
    \end{center}
\end{figure}

\begin{figure*}
    \begin{center}
    \includegraphics[scale=0.75,angle=-90]{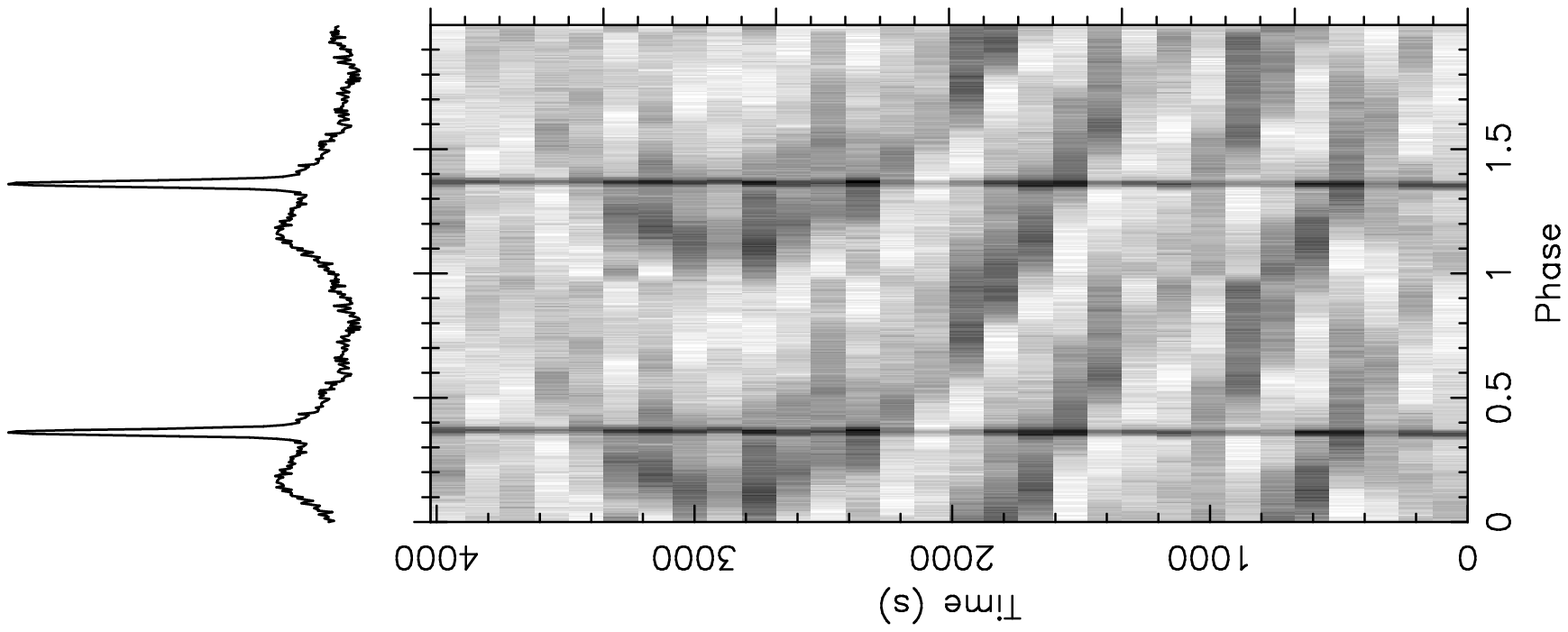}
    \hspace{2cm}
    \includegraphics[scale=0.75,angle=-90]{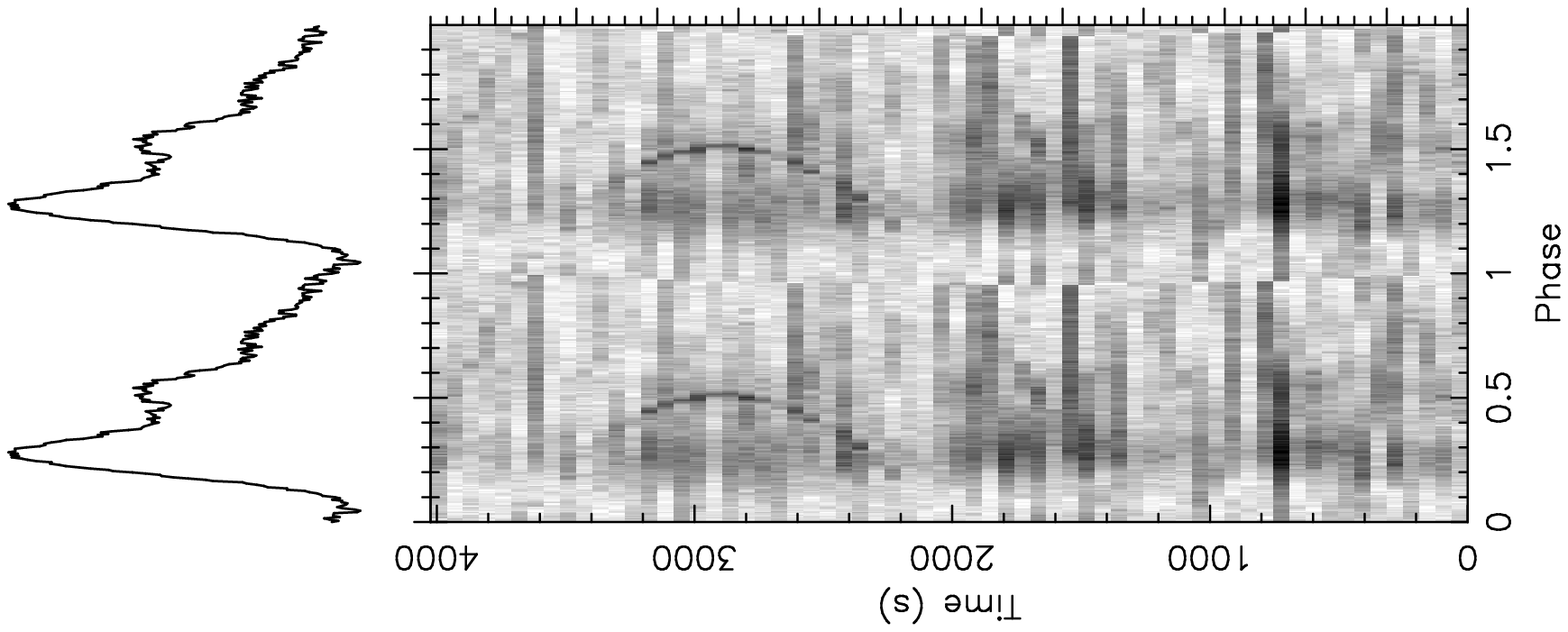}
    \vspace{1.7cm}
    \caption[]{Example {\sc presto} {\sc prepfold} folded profiles from the periodicity search at $4.85\,{\rm GHz}$ showing a detection of PSR~J1745$-$2900 and the anomalously drifting signal described in Section~\ref{ss:anom_bin} (left-hand panels). In the right-hand panels the same data are re-folded with an optimum $P,\,\dot{P}$ and $\ddot{P}$ to recover the anomalously drifting signal. The lower panels show folded subintegrations throughout the observation duration, while the upper panels show the folded and fully integrated pulse profile.}
    \label{Fig:1745_per}
    \end{center}
\end{figure*}

By using $T_{\rm sys}$, or the corresponding system equivalent flux
density $S_{\rm sys}=T_{\rm sys}/G\,({\rm Jy})$, as a useful marker of
the instantaneous sensitivity of our observing system, the effects
described above have been investigated. Values of $T_{\rm sys}$ at
three pertinent sky positions were measured: directly toward Sgr~A*
(labeled Sgr~A*~ON); at the same telescope elevation, $\alpha$, as
Sgr~A* $(\alpha<11^{\circ})$, but off source (Sgr~A*~OFF); and at high telescope
elevation $(\alpha >60^{\circ})$ close to the calibrator ${\rm NGC}\,7027$ but
also off source (${\rm NGC}\,7027$~OFF). Results of this analysis can
be seen in Figure~\ref{fig:cal_sens}, where the measurements of
$T_{\rm sys}$ at $4.85\,{\rm GHz}$ and $8.35\,{\rm GHz}$ over the
course of one year of observations is plotted. Average values of
$T_{\rm sys}$, and flux density limits, at all frequencies are also
given in Table~\ref{table:tsys_sefds_slims}. From simple differencing we
are able to investigate the relative contributions of $T_{\rm GC}$,
$T_{\rm elv}$ and $T_{\rm rec}$. At $4.85\,{\rm GHz}$, $T_{\rm GC}\sim
140\,{\rm K}$ and is the largest contributor to $T_{\rm sys}$,
whereas at $8.35\,{\rm GHz}$ $T_{\rm GC}\sim55\,{\rm K}$ which is less
than the combined effect of $T_{\rm elv}$ and $T_{\rm rec}$ at this
frequency.  In all of our measurements $T_{\rm atm}$ is degenerate
with values of $T_{\rm rec}$, however by comparison with values of the
system temperature measured at the zenith angle\footnote{\url{www.mpifr-bonn.mpg.de/effelsberg/astronomers}}, and corrected for
opacity effects, (denoted $T^{\dagger}_{\rm{rec}}$ in
Table~\ref{table:tsys_sefds_slims}) we can estimate the contribution from $T_{\rm atm}$ close to zenith. At $14.6\,{\rm
  GHz}$ $T_{\rm atm}\sim50\,{\rm K}$ which is the largest amongst our
calibrated values. At frequencies in the K-band $(18-27\,{\rm GHz})$ $T_{\rm atm}$ can vary by factors of a few 
depending upon the weather \citep{rtk+04}. Therefore, observations at $18.95\,{\rm GHz}$ 
were always performed in wintertime and under ideal weather conditions. $T_{\rm sys}$ in the Sgr~A* ON position was conservatively estimated 
using the measured value of $T_{\rm rec}=64\,{\rm K}$, $T_{\rm atm}\sim40\,{\rm K}$ - from the good 
weather conditions displayed in Figure 5. of \citet{rtk+04} and $T_{\rm GC}=20\,{\rm K}$ - by fitting a 
power law spectrum to the three lower frequency $T_{\rm GC}$ measurements (resulting spectral index of~$-1.4\pm0.3$).
The derived value of $T_{\rm sys}$ in the Sgr~A*~ON position comes to $124\,{\rm K}$. From test observations of PSR~B2020+28,
with known flux density at this frequency \citep{kxj+96}, and utilizing Equation~\ref{eq:radiometer}, we find agreement on 
this value of $T_{\rm sys}$ to within a factor of two.
  
At the two lowest frequencies, system temperature measurements are typically
stable for all telescope orientations and observing epochs over the
course of one year, with the exception of a period around MJD 57080 at
$8.35\,$GHz where a jump in $T_{\rm sys}$ is observed for all
telescope orientations. This might be attributed to adverse weather
conditions, or heightened levels of RFI on this day.
All sensitivity measurements with the PFFTS and XFFTS backends presented in this section 
were consistent to the ten per cent level with those from commensal observations using the PSRIX backend \citep{dep+18,lkg+16}.

\section{Results} 
\label{s:results}

\begin{figure*}
    \begin{center}
    \includegraphics[scale=0.75]{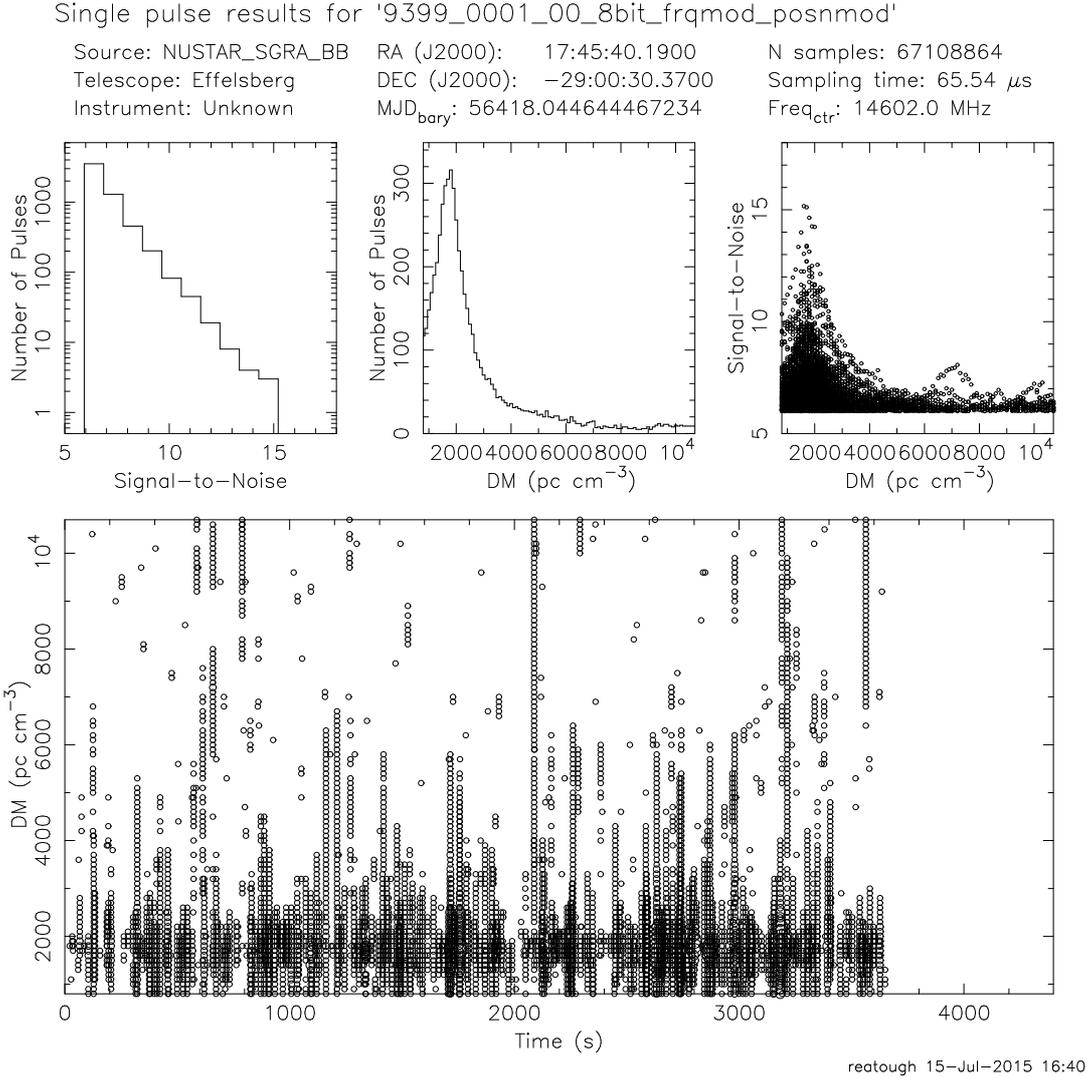}
    \caption[]{Example {\sc presto} {\sc single\_pulse\_search.py} diagnostic plot from our transient search showing a detection of 
single pulses from PSR~J1745$-$2900. Starting at the top left-hand panel and moving clockwise, the candidate plot shows; a histogram number of pulses as a function of pulse detection ${\rm S/N}$; a histogram of the number of pulses as a function of dispersion measure; a scatter diagram of the pulse ${\rm S/N}$ as a function of dispersion measure; a scatter diagram showing pulses as a function of dispersion measure and time through the observation - the size of a point is indicative of its ${\rm S/N}$. At the top, various observational diagnostics are given in plain text. } 
    \label{Fig:1745_sp}
    \end{center}
\end{figure*}

At the time of writing, no undiscovered pulsars or transients have been detected. Scatter diagrams showing example results from the periodicity search of individual epochs at each observing frequency can be seen in Figure~\ref{Fig:scatter_diag}. Such scatter diagrams were used to quickly select and display candidate detection diagnostics ({\sc presto prepfold} plots) of any promising pulsar candidates. For this task the so-called {\sc combustible-lemon} software tool was used\footnote{\url{https://github.com/ewanbarr/combustible-lemon}}. Typically up to $\sim 19,000$ candidates could be generated for each observing epoch. As is typical in these scatter diagrams, repeated detections of the same periodicity in independent observations, or data segments, are visible as ``columns" \citep{fsk+04,kel+09}. Whereas these can normally be attributed to terrestrial RFI detected at multiple sky locations our segmented acceleration search algorithm, applied to a single position on the sky, could also create detection ``columns" for genuine pulsar signals. 

Despite masking of its fundamental spin frequency and the first 32 integer harmonics, PSR~J1745$-$2900 was found in the periodicity search results of multiple epochs at $4.85$ and $8.35\,{\rm GHz}$; typically via fractional harmonics of the fundamental spin frequency (see Figure~\ref{Fig:scatter_diag}, upper panel). The upper panel of Figure~\ref{Fig:scatter_diag} shows $4.85\,{\rm GHz}$ results data from February $9^{\rm th}$ 2016 where two detections of PSR~J1745$-$2900 can be seen at the fundamental spin period in addition to multiple detection columns at shorter periods due to RFI. The corresponding {\sc presto prepfold} pulse profile is shown in the left hand panel of Figure~\ref{Fig:1745_per}. Further detection columns at $14.6$ and $18.95\,{\rm GHz}$ are also thought to be caused by RFI. Detections of PSR~J1745$-$2900 at the fundamental period are likely because the spin frequency decreased from its earlier value by $\sim2\times10^{-4}\,{\rm Hz}$ due to magnetic braking, and was no longer masked by the original ``birdie'' filters. PSR~J1745$-$2900 has also been detected in single pulse searches for transients signals in the majority of epochs (see Figure~\ref{Fig:1745_sp} for an example of a diagnostic plot of single pulse search results at $14.6\,{\rm GHz}$). Using Equation~1. in \citet{kakl+15}, and system sensitivity parameters outlined in Table~\ref{table:tsys_sefds_slims}, we estimate $6\sigma$ on source sensitivity limits to representative $1\,{\rm ms}$ duration single pulses of $1.4$, $0.6$, $1.0$ and $0.2\,{\rm Jy}$ at $4.85$, $8.35$, $14.60$ and $18.95\,{\rm GHz}$ respectively.     

Whereas detections of PSR~J1745$-$2900 were a useful validation of the data processing pipeline, their abundance either through fractional harmonics or transient bursts has slowed and potentially hampered the detection of other, possibly weaker, GC pulsars. In the following two subsections the detection of signals that are expected to be caused by either RFI or observational artefacts are described. While these signals are likely due to man-made effects, we detail them in order to inform future pulsar searches of the GC at these high observing frequencies.

\subsection{An anomalous repeating signal with a period of $3.74\,{\rm s}$}
\label{ss:anom_bin}
In the left-hand panels of Figure~\ref{Fig:1745_per}, which shows sub-integrations folded at the spin period, $P$, of the candidate in question (in this case PSR~J1745$-$2900 itself), a quadratically varying signal with a similar period is also visible. Such quadratic phase variation implies a signal that has a constant period derivative, $\dot{P}$, and is similar to the characteristics of a pulsar undergoing constant acceleration. This signal was first detected unambiguously in July 2014 and is detected in about 70 per cent of observations at $4.85\,{\rm GHz}$ thereafter. Re-folding the data with $P$, $\dot{P}$ and $\ddot{P}$ found with {\sc presto prepfold} the corrected pulse profile and subintegrations are visible in the right-hand panels of Figure~\ref{Fig:1745_per}. In this figure one can also see how the detection of PSR~J1745$-$2900 becomes smeared out relative to this signal. Over the data span presented here, we find an average barycentric spin period, with highly stochastic variations, of $3739^{+5}_{-9}\,{\rm ms}$.  The period of PSR~J1745$-$2900 of $\sim3763.5\,{\rm ms}$ at this time is surprisingly close \citep{efk+13}. 

No significant detection of this signal has been made in observations on identical azimuth and elevation tracks as Sgr~A*, but when the Sgr~A region had set. The latter could be because of reduced observing cadence and demonstrates the advantages of dual or multi-beam receivers that were unavailable in this work. 
While the trial DM at which the signal peaks in intensity is $0\,{\rm cm}^{-3}\,{\rm pc}$ the DM is uncertain because at these frequencies the dispersion delay across the band at $4.85$ and $8.35\,{\rm GHz}$ for an example DM of $1000\,{\rm cm}^{-3}\,{\rm pc}$ is only $37$ and $7\,{\rm ms}$ respectively; or one hundredth and five hundredths of the pulse width respectively. We note that similar candidate signals with a large pulse width are described in \citet{mq10} and also could not be ruled out as terrestrial due to a lack of measurable dispersion effects. Analysis of the polarization properties of the signal has revealed no clear Faraday rotation. In addition to the unusual pulse profile with extremely large duty cycle, these features indicate a terrestrial origin, perhaps due to airport radar or artefacts caused by observations of the bright GC region. Further observations will help to fully describe this signal and its origin.

\begin{figure}
    \begin{center}
    \includegraphics[height=\columnwidth,angle=90]{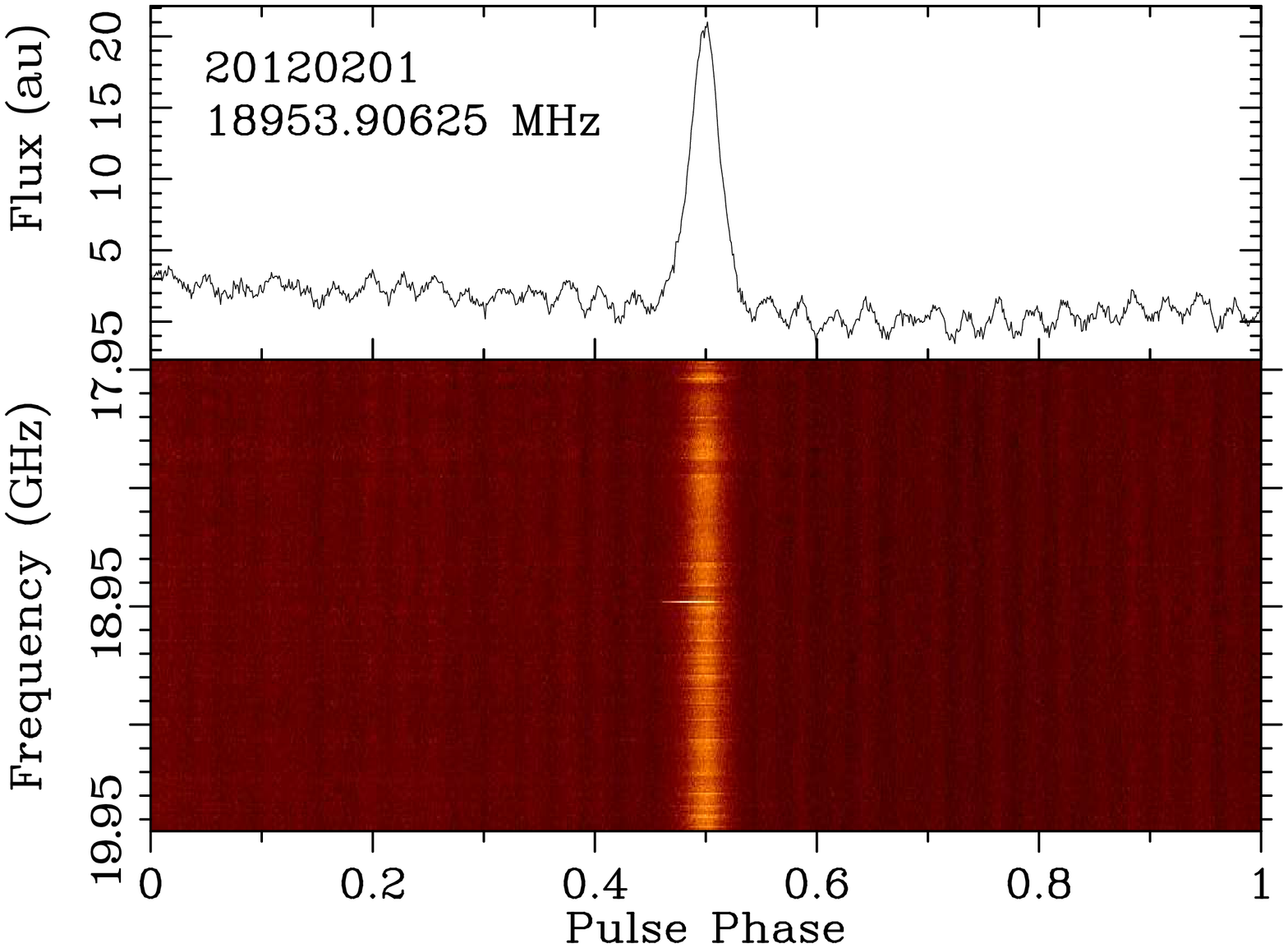}
    \caption{A $10\,{\rm s}$ duration dynamic spectrum centred around the $0.15\,{\rm Jy}$ single pulse event detected at $18.95\,{\rm GHz}$ on February $1^{\rm st}\,2012$. The top panel shows the integrated pulse profile (in arbitrary units) as a function of pulse phase. The lower panel shows frequency  subbands, also as a function of pulse phase, across $2\,{\rm GHz}$ around the central observing frequency of $18.95\,{\rm GHz}$. The``ripple-like" instabilities around the main pulse are discussed in Section~\ref{ss:transient}.} 
    \label{Fig:Kband_sp}
    \end{center}
\end{figure}

\begin{figure}
    \begin{center}
    \begin{minipage}{\linewidth}
    \includegraphics[height=\columnwidth,angle=-90]{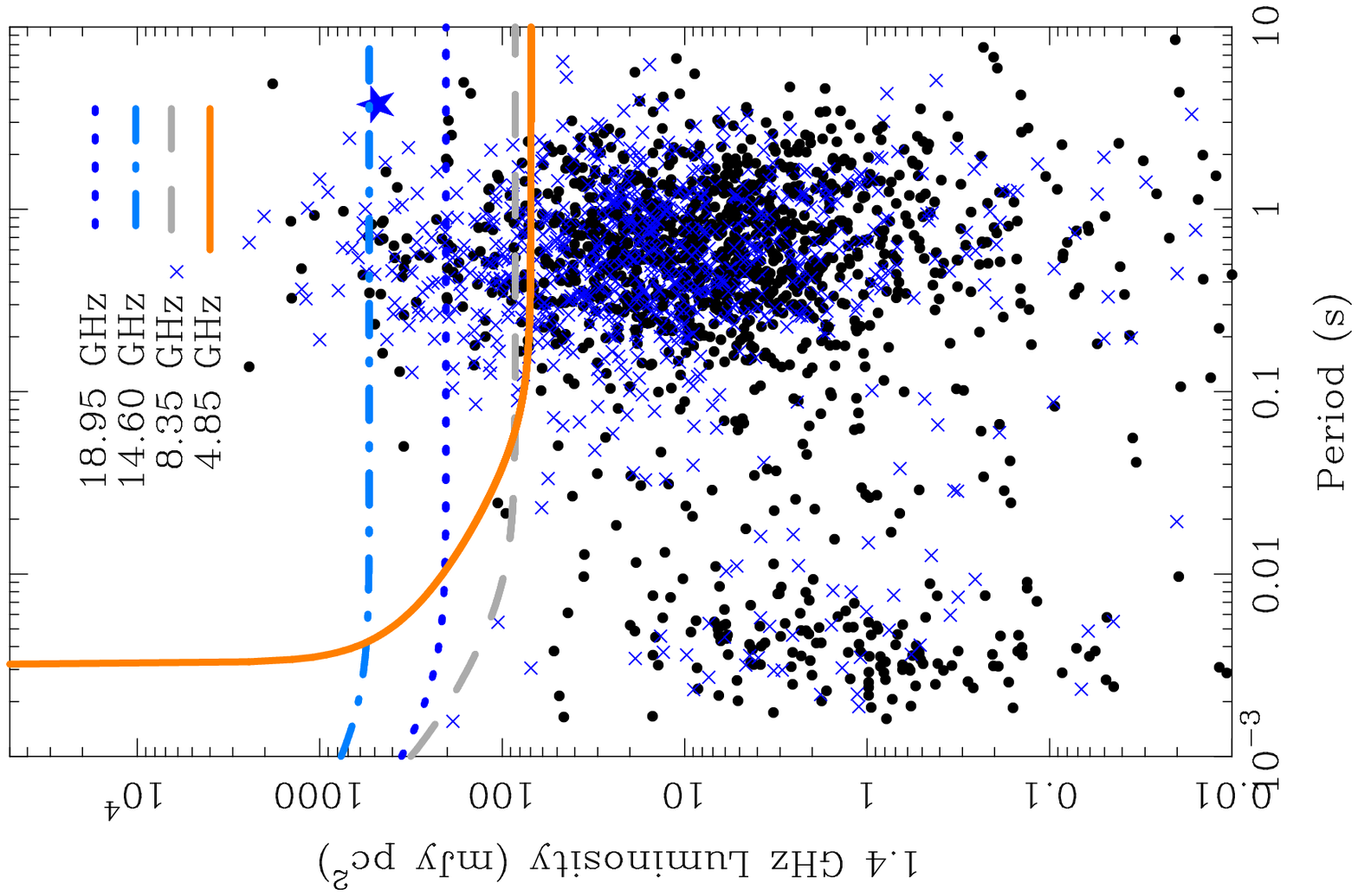}
    \caption{Luminosity at $1.4\,{\rm GHz}$ versus spin period of the known pulsar population (black dots and blue crosses) with luminosity information from the ATNF pulsar catalogue \citep{man05}. Blue crosses indicate pulsars with a measured spectral index. PSR~J1745$-$2900 is indicated by a blue star and its $1.4\,{\rm GHz}$ luminosity was derived using the measured spectral index from multi-frequency observations \citep{tek+15}. The sensitivity limits of the periodicity searches of our observations of the GC are over plotted with various thick colored lines that are scaled assuming an average pulsar spectral index of $-1.6$, a GC distance of $8.178\,{\rm kpc}$ \citep{grvty+III} and an average pulse width of $0.05P$. Only luminosity limits for  the full observation duration are plotted. The segmentation strategy of the acceleration search decreases sensitivity by a factor of $\sqrt{2}$ upon each consecutive segmentation (thereby raising the lines). Pulse broadening due to scattering follows published measurements \citep{spi14} and intra-channel dispersion smearing is for a single channel at the central observing frequency assuming a GC DM of $1778\,{\rm cm}^{-3}\,{\rm pc}$, based on observations of PSR~J1745$-$2900 \citep{efk+13}.} 
    \label{f:lum_sens}
    \end{minipage}
    \end{center}
\end{figure}

\subsection{A bright transient event at $18.95\,{\rm GHz}$}
\label{ss:transient}
Single pulse searches of an observation of Sgr~A* at $18.95\,{\rm GHz}$ on February $1^{\rm st}\,2012$, revealed a bright ${\rm (S/N\sim85)}$ broadband single pulse event of duration (or pulse width) $\sim0.5\,{\rm s}$ on barycentric MJD 55958.3462226  (Figure~\ref{Fig:Kband_sp}). We estimate a flux density of $0.15\,{\rm Jy}$ with a $6\sigma$ sensitivity of $0.01\,{\rm Jy}$ for pulses of this width. No DM could be measured due to the small sweep across the band at this frequency ($4.4\,{\rm ms}$ across the $2\,{\rm GHz}$ band at a DM equivalent to that of PSR~J1745$-$2900 of $1778\,{\rm cm^{-3}}\,{\rm pc}$) and broad pulse profile. Similar individual bright events were also identified in three further Sgr~A* observing epochs on barycentric MJDs 55959.2934368, 55968.3290126 and 55988.2245348. Accurate measurements of the relative arrival time of the single pulse events was not possible with the XFFTs backend because it was not connected to the observatory clock. Because of the lack of dispersion smearing as an astrophysical discriminator, we conducted an observation following the same azimuth and elevation track as Sgr~A* but at an alternative hour angle. During this observation another single pulse event with similar characteristics was detected on MJD~56030.2241117. This suggested possible source of terrestrial interference, perhaps due to a satellite uplink or downlink which are known to operate in this frequency range. We also note ``ripple-like'' instabilities around main pulse, possibly suggesting a satellite passing through the primary beam pattern. The possibility of emission from PSR~J1745$-$2900 before its first detection on April~$28^{\rm th}\,2013$ cannot be ruled out, however such broad individual single pulses have not been observed from this pulsar in subsequent observations \citep{efk+13,dep+18}.

\begin{table*}
  \centering
  \caption[]{The system temperature, $T_{\rm sys}$, quoted in the various published GC pulsar search analyses. The observing frequency in GHz is indicated in brackets. In all cases both the receiver, GC background and sky temperatures are included in $T_{\rm sys}$. Publications marked with a ${\clubsuit}$ assume a GC background temperature scaled from continuum surveys and those marked with a $*$ performed flux calibration via a noise diode. The $\spadesuit$ symbol denotes that in \citet{den09} the search was conducted at $2\,{\rm GHz}$, and $T_{\rm sys}$ given at other frequencies was for flux estimates of the discovered pulsars. In \citet{john06} the system equivalent flux density in units of Jansky has been converted to $T_{\rm sys}$, in units of Kelvin, by use of telescope gain values given in the webpage at the bottom of this table$^{11}$. Note. errors are not given and frequencies have been rounded to one decimal place for clarity.      
  }
  \label{table:tsys_surv}
  \begin{tabular}{llcccccccccccc}
\hline
   
   \multicolumn{1}{c}{Publication} & \multicolumn{1}{c}{Tel.}	&  $T_{\rm sys}\,({\rm K})$ &  $T_{\rm sys}\,({\rm K})$ &  $T_{\rm sys}\,({\rm K})$ & \ditto & \ditto & \ditto & \ditto & \ditto & \ditto & \ditto & \ditto  \\
   && $1.4\,{\rm GHz}$ & $2.0\,{\rm GHz}$ & $3.1\,{\rm GHz}$ & $4.8\,{\ditto}$ & $4.9\,{\ditto}$ & $8.4\,{\ditto}$ & $9.0\,{\ditto}$ & $14.4\,{\ditto}$ & $14.6\,{\ditto}$ & $14.8\,{\ditto}$ & $19.0\,{\ditto}$\\
\hline
\hline
   \citet{john06}$^{\clubsuit}$ 			& Pks. & $-$ & $-$ & $541$ & $-$ & $-$ & $123$ & $-$ & $-$ & $-$ & $-$ & $-$ \\
   \citet{den09}$^{\clubsuit}$ 			& GBT & $32$ & $27\spadesuit$ & $-$ & $19$ & $-$ & $-$ & $27$ & $-$ & $-$ & $-$ & $-$	\\
   \citet{mq10}*         	& GBT  & $-$ & $-$ & $-$ & $-$ & $-$ & $-$ & $-$ & $35$ & $-$ & $38$ & $-$	\\
   This work*			    & Eff.  & $-$ & $-$ & $-$ & $-$ & $200$ & $126$ & $-$ & $-$ & $194$ & $-$ & $79$ \\
\hline
\end{tabular}
$^{11}\;$\footnotesize{\url{https://www.parkes.atnf.csiro.au/observing/documentation/users_guide/html/pkug.html\#Receiver-Fleet}}
\end{table*}

\subsection{Recovered fractions of a Galactic Centre pulsar population}
\label{ss:rec_frac}
Figure~\ref{f:lum_sens} shows the $1.4\,{\rm GHz}$ ``pseudo-luminosity'' (hereafter termed ``luminosity''; given by $L_{1400}=S_{1400}d^2$ where $S_{1400}$ is the $1.4\,{\rm GHz}$ flux density and $d$ is the best known distance) of 2125 known pulsars as a function of spin period 
with information from the ATNF pulsar catalogue version $1.62$ \citep{man05}. All pulsars in this sample have flux density measurements at either $1.4\,{\rm GHz}$ or $0.4\,{\rm GHz}$ (with the exception of PSR~J1745$-$2900, see below). For those pulsars with no flux density information at $1.4\,{\rm GHz}$ ($11\,{\rm per\,cent}$ of the total sample), we extrapolated from $0.4\,{\rm GHz}$ to this frequency using either the known spectral index (pulsars marked with blue crosses have a measured spectral index) or assuming a recently derived average pulsar spectral index of $-1.6$ \citep{jvsk+18}. For PSR~J1745$-$2900 the $1.4\,{\rm GHz}$ flux density was extrapolated from the measurements described in \citet{tek+15}. The $10\sigma$ luminosity sensitivity limits of periodicity search observations presented in this work are marked with various colored and dashed/dotted lines. Each limit is scaled to its value at our chosen ``reference frequency'' of $1.4\,{\rm GHz}$ also assuming an average pulsar spectral index of $-1.6$ and according to the GC distance. Therefore pulsars that lie above a given limit would be detected with at least $10\sigma$ significance if placed at the GC distance. 

The single reference frequency of $1.4\,{\rm GHz}$ is useful for comparing the approximate relative sensitivity of our multi-frequency search observations and for determining what fraction of a hypothetical GC pulsar population could be detected \citep{ckl+06}. For instance, assuming the GC pulsar population follows the same luminosity distribution as the currently known pulsars, have an intrinsic pulse width of $0.05P$ and neglecting any binary motion, we determine that approximately 11, 10, one and four per cent of this pulsar population would be detected at  $4.85,\,8.35,\,14.6,\,{\rm and}\,18.95\,{\rm GHz}$ respectively. Assuming a larger intrinsic pulse width of $0.1P$ the detected percentages are reduced marginally to eight, seven, one and three per cent at $4.85,\,8.35,\,14.6,\,{\rm and}\,18.95\,{\rm GHz}$ respectively.

Differences in the recovered population fraction can occur when scaling pulsar flux densities to alternative reference frequencies (particularly at the two highest frequencies of $14.6$ and $18.95\,{\rm GHz}$). Because our observations cover a wide range in observing frequency, we have created and analysed the equivalent period luminosity diagrams at each frequency independently. In these analyses frequency scaling of the survey luminosity limits is not required. Also, to address the physical property of the observed spread in pulsar spectral indices, we have investigated the effects of choosing not just a single average spectral index (for those pulsars with no spectral index information) but a random selection from a Gaussian distribution with an average spectral index of $-1.6$ and a standard deviation of $0.54$, as given in \citet{jvsk+18}. These results (and those at the reference frequency of $1.4\,{\rm GHz}$) are presented in Appendix~\ref{s:appendixA} in Table~\ref{t:rec_frac} and can be seen in Figures~\ref{f:lum_sens_app1} and \ref{f:lum_sens_app2}. The effects of possible broken spectral power laws have not been taken into account and the numbers reported should be treated as approximate upper limits. We note that despite accounting for both effects described above, in all cases the recovered fraction of pulsars from this hypothetical population is no better than 13 per cent, illustrating that the intrinsic sensitivity of these observations to typical pulsars at the GC is still markedly low. The reductions in $T_{\rm obs}$ applied in our acceleration search for compact binary systems compound this sensitivity problem further. Similar population detection analyses have recently been given in the results from GC pulsar searches conducted at millimetre wavelengths \citep{tde+21,lde+21}.

In addition, \citet{lbh+15} have shown that the sensitivity of pulsar searches in the Arecibo pulsar ALFA (PALFA) survey at $1.4\,{\rm GHz}$ degrades (by up to a factor of 10) for spin periods above $100\,{\rm ms}$ due to red noise and RFI effects present in their data. To investigate if such effects are observed in the data used in this work $-$ where in particular atmospheric fluctuations may produce red noise features $-$ we have injected simulated pulsar signals into example data sets at each observing frequency and measured any reductions in sensitivity. Evidence for a radio frequency dependence in the detrimental effects on sensitivity has been found, which we would indeed expect for red noise dominated by atmospheric effects which worsen towards K-band (see Figure~\ref{f:red_noise_eff}). This analysis is described in detail in Appendix~\ref{s:appendixB} along with the updated sensitivity, accounting for red noise effects for spin periods $>0.1\,{\rm s}$, plotted in Figures~\ref{f:lum_sens_app1} and \ref{f:lum_sens_app2}. The recovered fraction of pulsars are further reduced by a few percentage points due to these effects and is indicated in Table~\ref{t:rec_frac} by the numbers in parentheses.

\section{Discussion}
During the course of this work a number of aspects regarding the efficacy of searches for pulsars in the GC have been identified. 
In the following subsections we discuss two more areas of importance and finish by looking at some of the future prospects for GC pulsar searches.   

\subsection{Background sky brightness temperature toward the GC}
 As noted in \citet{john06} and \citet{mq10}, and as shown in Table~\ref{table:tsys_sefds_slims}, the GC background brightness temperature, $T_{\rm GC}$, can have a significant effect on the sensitivity of GC pulsar searches. In Table~\ref{table:tsys_surv} the various measurements, or estimates, of $T_{\rm sys}$ whilst on source (therefore including $T_{\rm GC}$) in published GC pulsar searches have been collected and placed with the calibrated measurements from this work. There is a large discrepancy between the figures presented in observations performed at the lowest frequencies $\nu<3.1\,{\rm GHz}$ in \citet{john06} and \citet{den09}. From our measurements outlined in Table~\ref{table:tsys_sefds_slims}, and the discussion in \citet{john06}, we expect $T_{\rm GC}$ to dominate $T_{\rm sys}$ for frequencies $\lesssim8\,{\rm GHz}$. For example we already find $T_{\rm GC}=137(16)\,{\rm K}$ at the frequency of $4.85\,{\rm GHz}$ $-$ in tension with the estimates of $T_{\rm sys}$ of $19\,{\rm K}$ at $4.8\,{\rm GHz}$ in \citet{den09}.   
 At higher frequencies of $8.4\,{\rm GHz}$ our measurements are consistent with those estimates presented in \citet{john06}. At frequency $14.6\,{\rm GHz}$ our measurement appears to be in conflict with that given at $14.8\,{\rm GHz}$ in \cite{mq10}. We postulate that the sensitivity at Effelsberg might have been degraded due to both the extremely low elevation of Sgr~A* (although measurements of $T_{\rm sys}$ at $14.6\,{\rm GHz}$ in the Sgr~A* ON and Sgr~A* OFF positions suggest this is not the case $-$ see Table~\ref{table:tsys_sefds_slims}), poor atmospheric conditions during the single calibration observation at this frequency and the higher receiver temperature. Repeated measurements at this frequency would have been beneficial in resolving this discrepancy.    

 The addition of our measurements to existing published figures highlights some inconsistencies and also an overall sparsity of directly calibrated sensitivity measurements in GC pulsar searches.  
While continuum measurements of the GC region with single dish telescopes (e.g.~\citealt{rfr+90,lzc+08}) have been beneficial for estimating the sensitivity of pulsar searches, these cannot take into account weather or instrumental effects during the pulsar search.

\subsection{Binary search considerations for pulsars closely orbiting Sgr~A*} 
\label{ss:binarry_search_con}
\begin{figure}
    \begin{center}
        \includegraphics[height=\columnwidth,angle=-90]{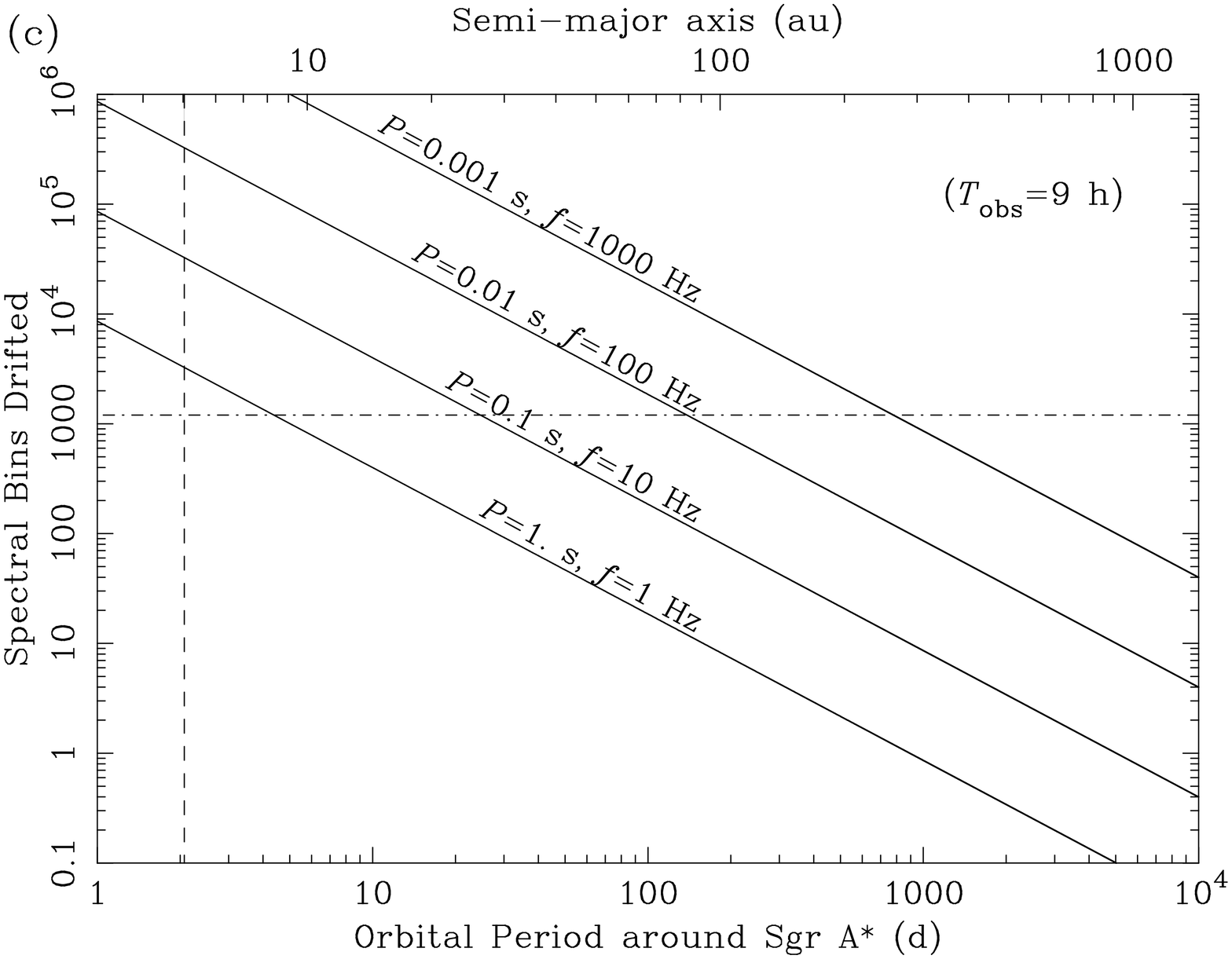} 
        \includegraphics[height=\columnwidth,angle=-90]{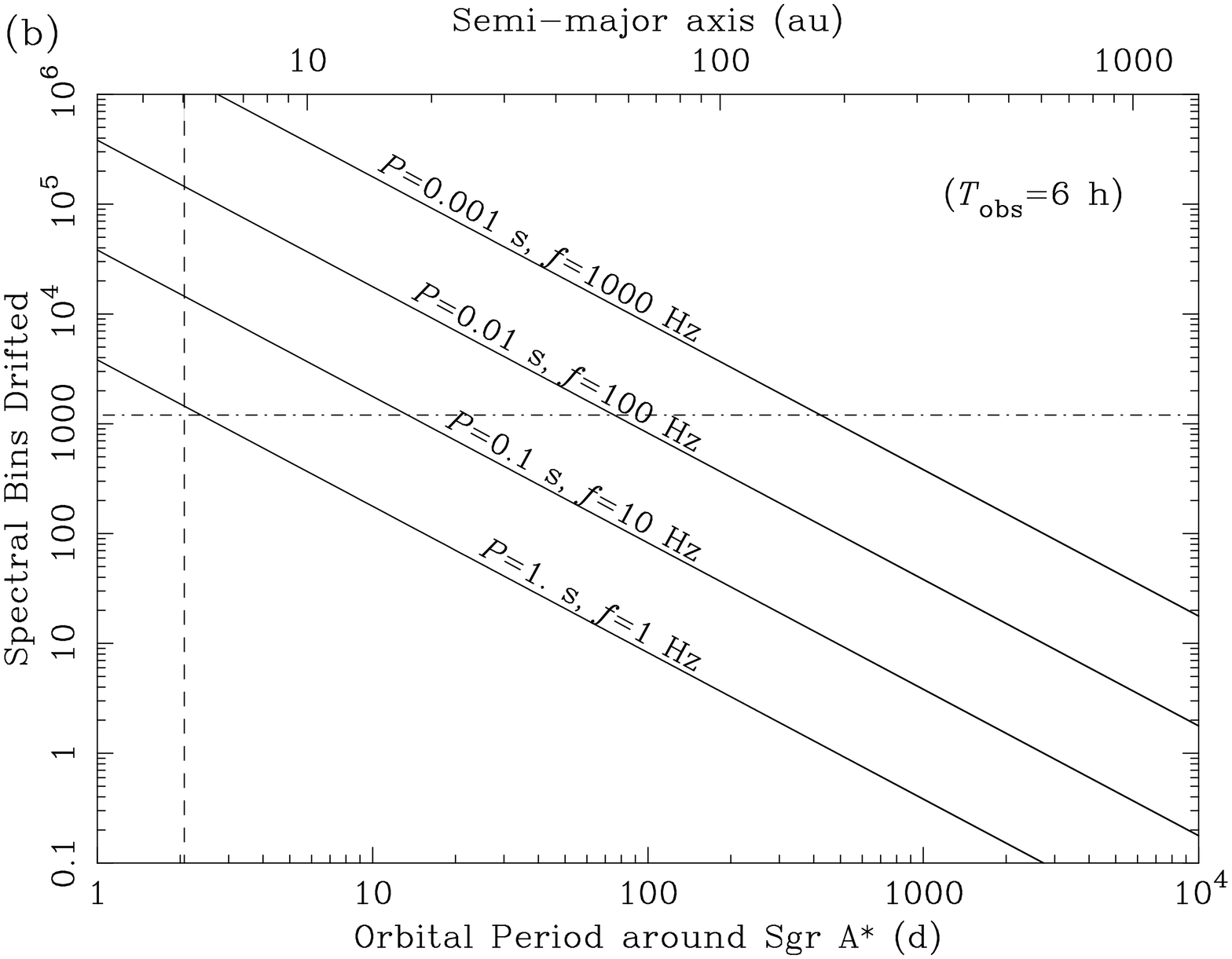}
        \includegraphics[height=\columnwidth,angle=-90]{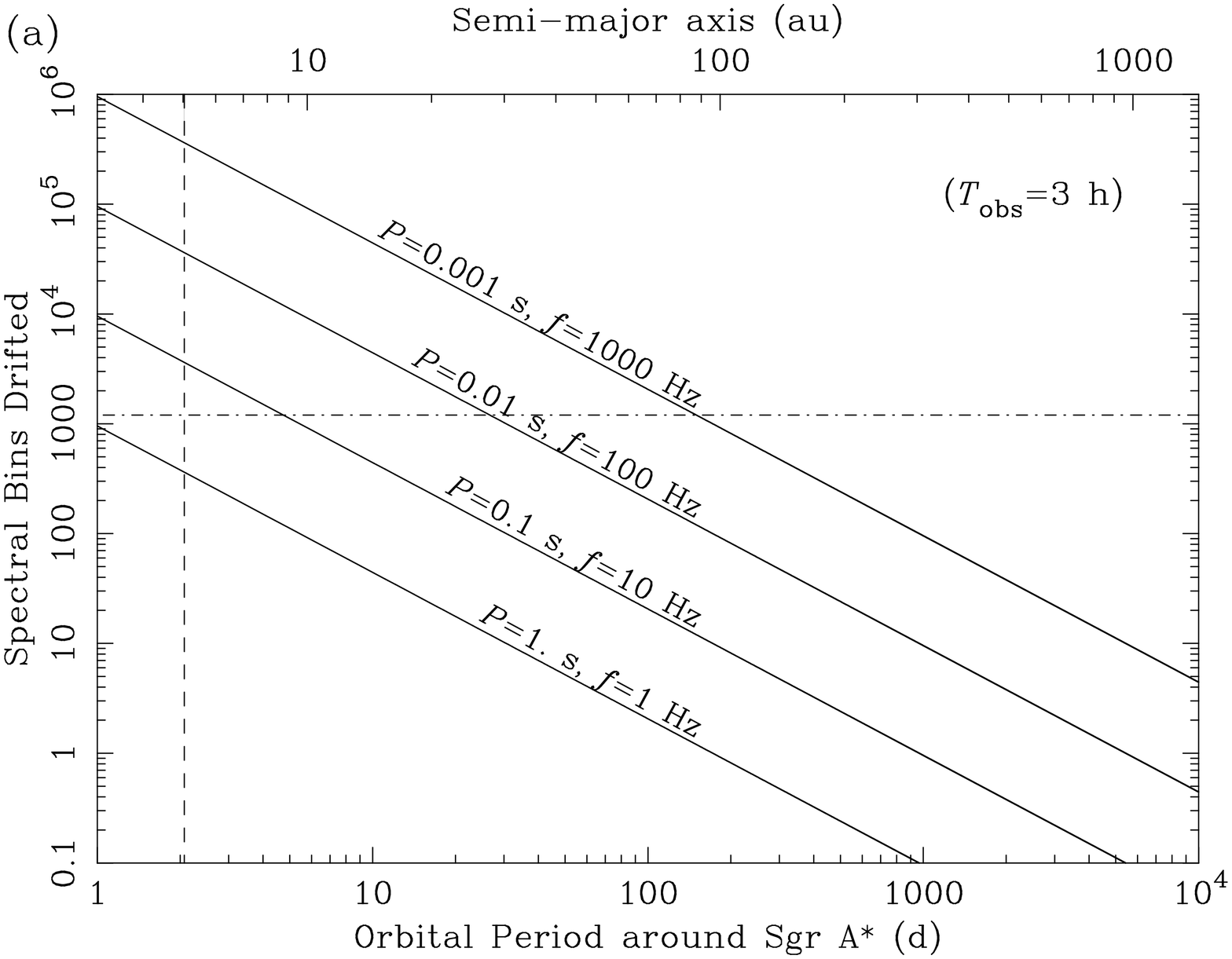}
        \caption[]{The theoretical number of spectral bins the spin frequency $f$ (or period $P$) drifts in the Fourier spectrum for pulsars in circular orbits around Sgr~A*, for three representative observation lengths of $3$, $6$ and $9\,{\rm h}$ (panels a, b and c respectively) and which occur at the phase of the orbit where the l.o.s. acceleration is at a maximum. For the orbital period and observation length ranges plotted, the l.o.s. acceleration can be assumed to be approximately constant. The dot dashed horizontal line indicates the $z_{\rm max}$ value used in this work of 1200 and the vertical dashed line shows the orbital period for a pulsar in an orbit with a merger time of $10^{7}\,{\rm y}$ $-$ the typical lifetime of a normal pulsar.}
    \label{f:sgra_accn_ndrift}
    \end{center}
\end{figure}

For searches of orbiting pulsars, Sgr~A* presents a unique set of conditions because of its extreme mass. Even pulsars in long period orbits $(P_{\rm b}\lesssim800\,{\rm d})$ can undergo sufficiently large acceleration that the methods used in this work which search for signals with constant spin frequency drift in the Fourier spectrum (viz. {\sc presto accelsearch}) can be overcome. In Figure~\ref{f:sgra_accn_ndrift} we plot the number of spectral bins drifted in the Fourier spectrum as a function of orbital period around Sgr~A* for pulsars in circular orbits, for a number of representative spin periods and observation lengths (Figure~\ref{f:sgra_accn_ndrift} panels a, b and c). The intersection of diagonal lines with the horizontal dot dashed line (the current maximum value of $n_{\rm drift}$ searched given by the {\sc zmax} parameter in {\sc accelsearch}) gives the lower limit of the orbital period for a pulsar with fundamental, or harmonics, of that spin frequency that are detectable. Because $n_{\rm drift}\propto T_{\rm obs}^2$, the smearing of spectral features is much larger in the necessarily deep observations that might be required to detect pulsars at the GC distance ($T_{\rm obs}=6-9\,{\rm h}$, for $100\,{\rm m}$ class dishes). For example, if a maximum spin frequency of $1000\,{\rm Hz}$ is considered (a spin frequency that covers known MSPs and most of their harmonics) the minimum circular orbital periods around Sgr~A* detectable are: $P_{\rm b}\sim 780\,{\rm d}$, $a=0.3\,{\rm m\,s^{-2}}$ for $T_{\rm obs}=9\,{\rm h}$; $P_{\rm b}\sim 420\,{\rm d}$, $a=0.8\,{\rm m\,s^{-2}}$ for $T_{\rm obs}=6\,{\rm h}$ and $P_{\rm b}\sim 150\,{\rm d}$, $a=3.1\,{\rm m\,s^{-2}}$ for $T_{\rm obs}=3\,{\rm h}$. For longer spin periods the minimum orbital periods detectable are correspondingly shorter. 

\begin{figure}
    \begin{center}
        \includegraphics[height=\columnwidth,angle=-90]{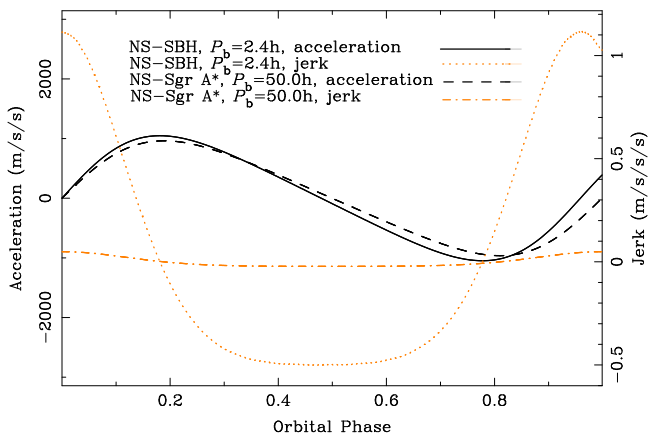}
        \caption[]{l.o.s acceleration and jerk as a
          function of orbital phase for two extreme pulsar black hole
          systems viewed edge on to the orbital plane. The solid black and orange dotted lines represent the acceleration and jerk (respectively) in a compact pulsar 
          stellar mass black hole binary (NS$-$SBH, $P_{\rm b}\,=\,2.4\,{\rm h}$, $e=0.1, m_{\rm c}=30\,{\rm M_{\odot}}$).
          The dashed black and dot-dashed orange lines mark the acceleration and jerk
          of a pulsar closely orbiting Sgr~A* (NS$-$Sgr~A*, $P_{\rm b}=50.0\,{\rm h}$, $e=0.1$, $m_{\rm c}=4\times10^6\,{\rm M_{\odot}}$).}
    \label{f:acc_jerk_sgra_comp}
    \end{center}
\end{figure}

For the purpose of investigating how robust binary pulsar searches around Sgr~A* are, a useful physical minimum orbital period to consider can be inferred by setting the time scale of coalescence due to the emission of gravitational waves, $\tau_{\rm GW}$, equal to the typical lifetime of a pulsar, $\tau_{\rm PSR}, $ where $\tau_{\rm PSR}\sim10^7\,{\rm y}$ (P.~Freire private comm. and see Appendix~A2.7 in \citealt{lorkra}). Such an exercise results in $P_{\rm b} \simeq 50\,{\rm h}$, average velocity $0.09\,c$ and a maximum l.o.s acceleration of $\simeq1000\,{\rm m\,s^{-2}}$ for systems viewed edge-on. Orbital accelerations of similar magnitude are currently only seen in extremely compact or compact and eccentric binary double neutron star (DNS) systems such as e.g. PSR~J0737$-$3039A/B and PSR~J1757$-$1854 \citep{kra06,cck+18}. The comparatively  longer orbital period of the hypothetical Sgr~A* pulsar described above means that higher order effects detrimental to binary pulsar searches  $-$ such as the rate of change of acceleration, known as ``jerk" $-$ are reduced. This effect is illustrated explicitly in Figure~\ref{f:acc_jerk_sgra_comp} where the acceleration and jerk of a pulsar throughout one orbit of a hypothetical compact $(P_{\rm b}=2.4\,{\rm h})$ pulsar stellar mass black hole, NS$-$SBH, and an extreme $(P_{\rm b}=50.0\,{\rm h})$ NS$-$Sgr~A* system  are plotted. The acceleration in both systems is approximately equivalent, peaking at around $1000\,{\rm m}\,{\rm s}^{-2}$ whereas the jerk exhibited by the NS$-$Sgr~A* is roughly 20 times smaller than that of the  NS$-$SBH. Using the commonly accepted limit for the application of acceleration searches that $T_{\rm obs} < P_{\rm b}/10$ \citep{ransom03,ng15}, higher order jerk effects in the Sgr~A* pulsars described above should only become relevant for observations where $T_{\rm obs}\gtrsim5\,{\rm h}$ for the the most extreme $P_{b} = 50\,{\rm h}$ system. 

Potential expansion of the {\sc zmax} term in {\sc presto accelsearch} specifically to deal with the demands of deep acceleration searches of the Sgr~A* region are currently under discussion (S.~Ransom private comm.). We also note that the most recent versions of {\sc presto accelsearch} can now search for and compensate higher order quadratic frequency drifts due to jerk \citep{ar+18}, potentially increasing the integration time that can be searched for binaries. For extremely long observations $(T_{\rm obs}\sim9\,{\rm h})$ time domain orbital template matching techniques will likely offer the highest sensitivity to a wide range of GC binary pulsars \citep{kek+13,akc+13}.

\subsection{Prospects for future pulsar searches of the GC}
The conditions in and toward the GC create a ``perfect storm" of observational requirements that are 
detrimental to searches for radio pulsars. In no particular order these are: its large distance ($\sim 84\,{\rm per\,cent}$ of known pulsars are closer than the GC due to their relative weakness as radio sources - \citealp{le+17}); high levels of pulse scattering (while scattering might be less than previously predicted it is still large enough to smear out MSPs at frequencies $\lesssim 7\,{\rm GHz}$ - \citealp{spi14}); intense background emission (the GC is the brightest region of the Galaxy at radio wavelengths reducing system sensitivity - Section~\ref{ss:sensitivity}); the typically steep spectrum of pulsar emission (at the high frequencies necessary to combat scattering pulsar emission is weaker); long observation times (the combined effects of steep pulsar spectra and the large GC distance currently necessitate long observations that can reduce sensitivity to binary pulsars or pulsars orbiting Sgr~A* - Section~\ref{ss:binarry_search_con}). 

To combat all of these effects simultaneously an interferometer with large collecting area and that operates at frequencies $\gtrsim 7\,{\rm GHz}$ is optimal. Interferometers offer the added benefit of resolving out part of the bright GC background that single dishes are exposed to. The relative weight of each of these effects and the optimal balance of observing parameters is not yet fully understood, however, pulsar searches using interferometers are already underway with Atacama Large Millimeter/submillimeter Array (ALMA)  \citep{lde+21} and the Karl G. Jansky Very Large Array (VLA) \citep{Wharton:thesis}. To fully rule out scattering effects in the GC, multi-epoch searches for flat spectrum pulsars have been conducted at millimetre wavelengths with the Institut de Radioastronomie Millim\'{e}trique (IRAM) $30\,{\rm m}$ telescope \citep{tde+21}. At Effelsberg, GC pulsar searches with nearly three times the sensitivity of those presented here (thanks to a new a broad-band $4-8\,{\rm GHz}$ receiver) will be given by Desvignes~et~al. (in prep.).

The instantaneous sensitivity for pulsar searches offered by next generation telescopes such as MeerKat \citep{sk+16,kmb+16}, Next Generation VLA \citep[ngVLA - ][]{bower_ngvla} and SKA1-mid \citep{eat15ska} might allow at least an order of magnitude improvement in survey depth and will also facilitate an increased binary parameter space that can be searched thanks to the reduction in the necessary integration length\footnote{Detection figures for GC pulsars shown in \citet{eat15ska} are given before the agreed ``re-baselining'' of SKA1-mid and are likely to be reduced. At MeerKat, the possibility of receivers operating $>2.5\,{\rm GHz}$ remains uncertain}.  

\section{Summary}
A multi-epoch survey for binary pulsars and fast transients in the Galactic Centre at frequencies $(4.85,\,8.35,\,14.60,\,18.95)\,{\rm GHz}$ was carried out between February 2012 and February 2016 using 
the \mbox{Effelsberg$\,$100-m} radio telescope. The various high radio frequencies utilized decrease the 
deleterious effects of strong pulse scattering that exist in this region. Comprehensive acceleration searches have been
conducted on progressively shorter segments of the full observation lengths to increase sensitivity to relativistic binary pulsars.
This survey is the first time that observations of the Galactic Centre have been regularly repeated over a time-span of the order of years; highly beneficial
for detecting pulsars undergoing relativistic binary motion or exhibiting transient phenomena. An anomalous repeating signal with a spin period close to that of PSR~J1745$-$2900 and a handful of single pulse events have been identified, but no previously unknown pulsars have been detected. Sensitivity measurements of our observing system have revealed that, at best, only 13 per cent of the known pulsar population could be detected by these searches if it were placed at the Galactic Centre distance. We also show that analysis of current or future deep observations $(9\,{\rm h})$ with existing pulsar search tools may struggle to detect some millisecond pulsars in orbits of less than $800\,{\rm d}$ around Sgr~A*.  Through accurate calibration of our observing system and investigation into previously unaccounted effects we demonstrate that earlier pulsar searches of the Galactic Centre may have overestimated their sensitivity. Future observatories with increased sensitivity, and the use of interferometers in periodicity searches, will improve the power to discover as yet undetected Galactic Centre pulsars.

\section*{Acknowledgements}

The authors acknowledge financial support by the European Research Council for the ERC Synergy Grant BlackHoleCam under contract no. 610058. This work was based on observations with the $100\,{\rm m}$ telescope of the Max-Planck-Institut 
f\"{u}r Radioastronomie at Effelsberg. RE is supported by a ``FAST Fellowship" under the ``Cultivation Project for FAST Scientific Payoff and Research Achievement of the Center for Astonomical Mega-Science, Chinese Academy of Sciences (CAMS-CAS)". 
PT was supported for this research through a stipend from the International
Max Planck Research School (IMPRS) for Astronomy and Astrophysics at the Universities of Bonn and Cologne. The authors wish to thank Lorenz Huedepohl and Ingeborg Weidl of the Max Planck Computing and Data Facility for their support with the \emph{Hydra} supercomputer. The authors also wish to thank Dr.~A.~Kraus and Dr.~A.~Jessner for observational assistance at Effelsberg.

\section*{DATA AVAILABILITY}
The data underlying this article will be shared on reasonable request to the corresponding author.


\bibliographystyle{mnras}

\newcommand{\noop}[1]{}


\appendix

\section{Alternative period luminosity diagrams}
\label{s:appendixA}
\begin{figure*}
    \begin{center}
    \includegraphics[height=6.5cm,angle=-90]{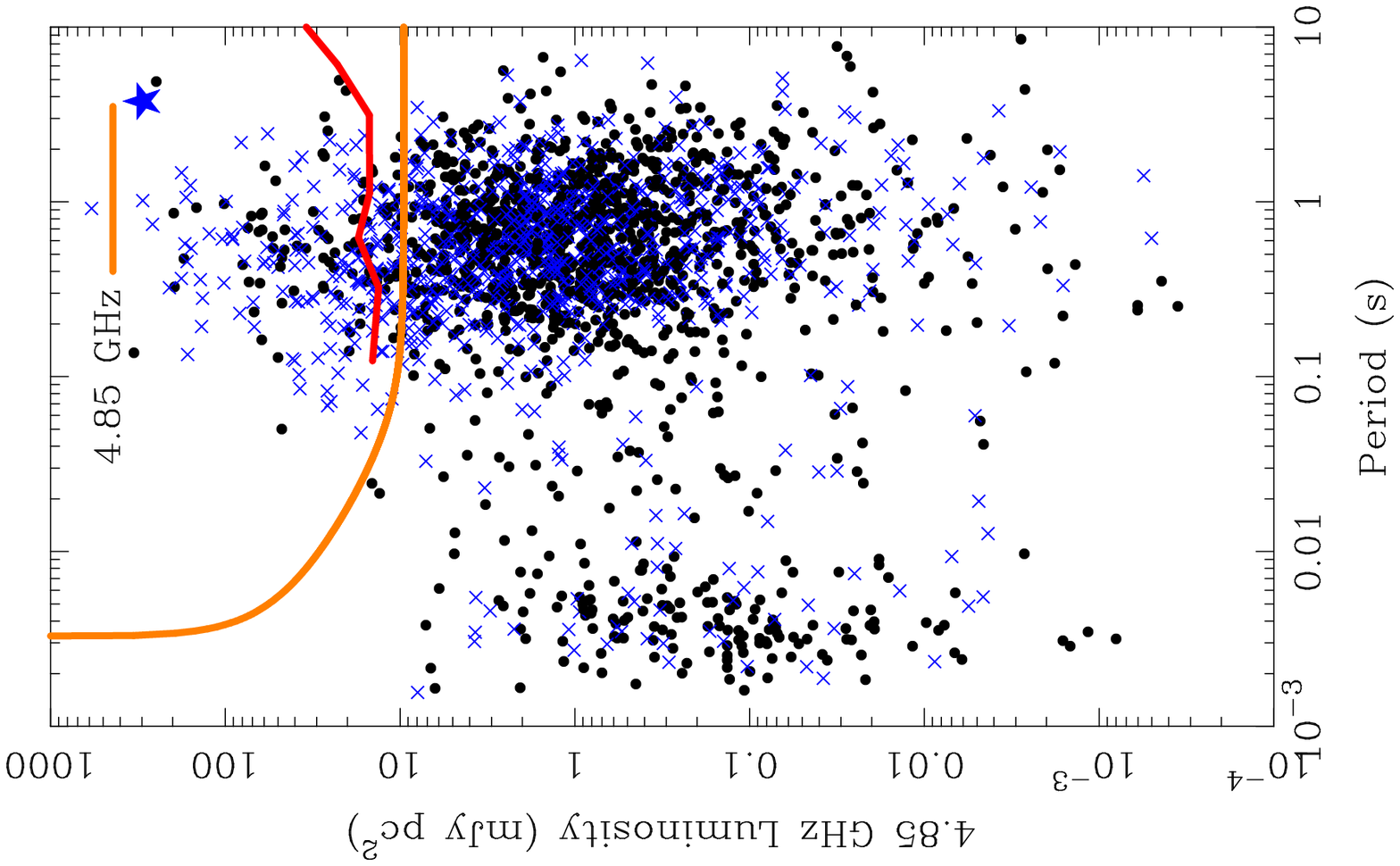}
    \hspace{1.2cm}
    \includegraphics[height=6.5cm,angle=-90]{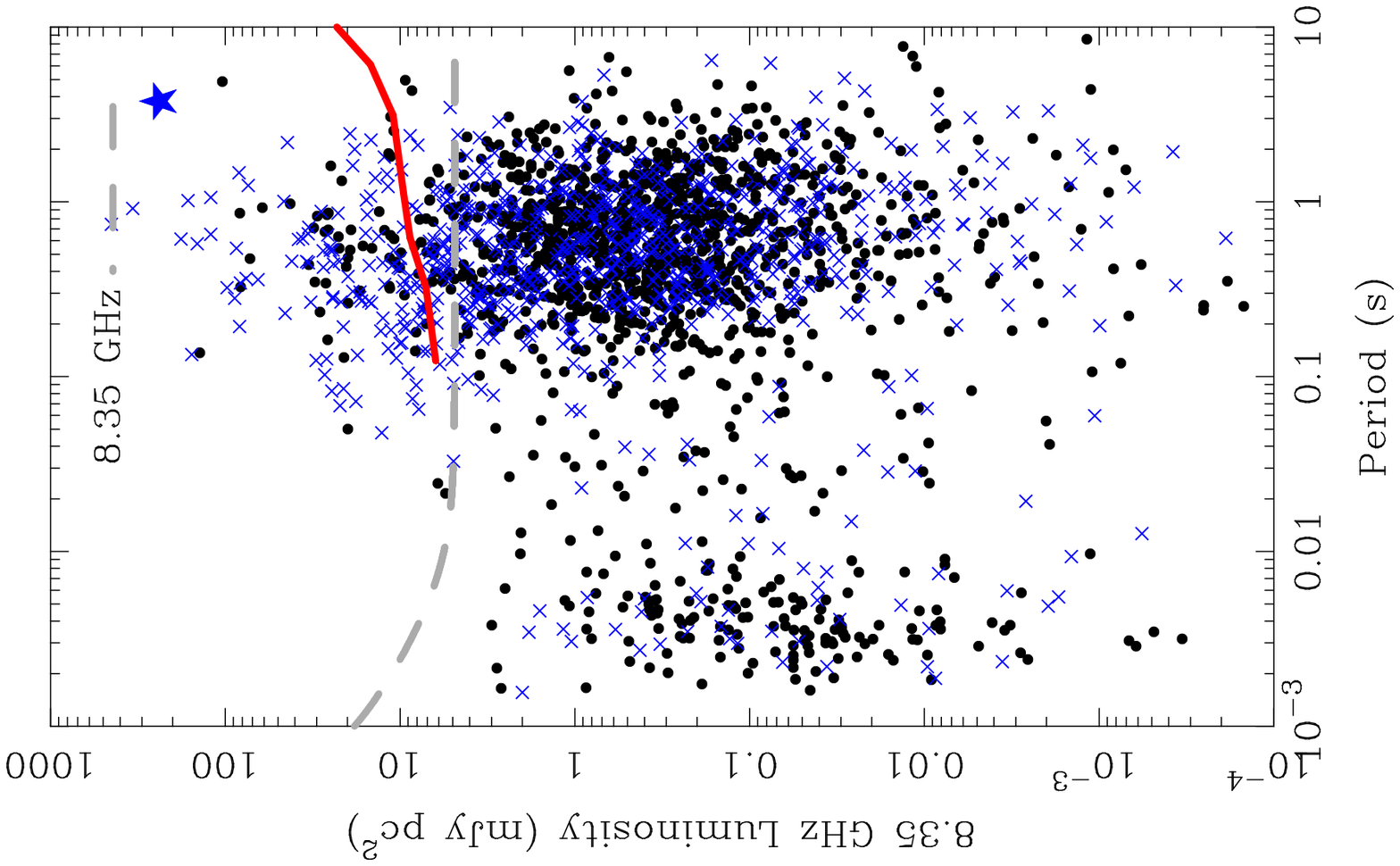}
    \caption{Luminosity versus spin period of the known pulsar population at $4.85\,{\rm GHz}$ (left hand panel) and $8.35\,{\rm GHz}$ (right hand panel). In these diagrams the pulsar luminosity is scaled from either $1.4\,{\rm GHz}$ and $0.4\,{\rm GHz}$ using the known spectral index (pulsars with an index are marked with blue crosses) or a spectra index of $-1.6$. The estimate of the $10\sigma$ sensitivity limit of our GC search observation (for the maximum observing duration) is marked with a line. The red line indicates measured $10\sigma$ sensitivity limits from the injection of simulated pulsar signals which account for red noise effects. See Appendix~\ref{s:appendixB} for more details about this limit.}  
    \label{f:lum_sens_app1}
    \end{center}
\end{figure*}

\begin{figure*}
    \begin{center}
    \includegraphics[height=6.5cm,angle=-90]{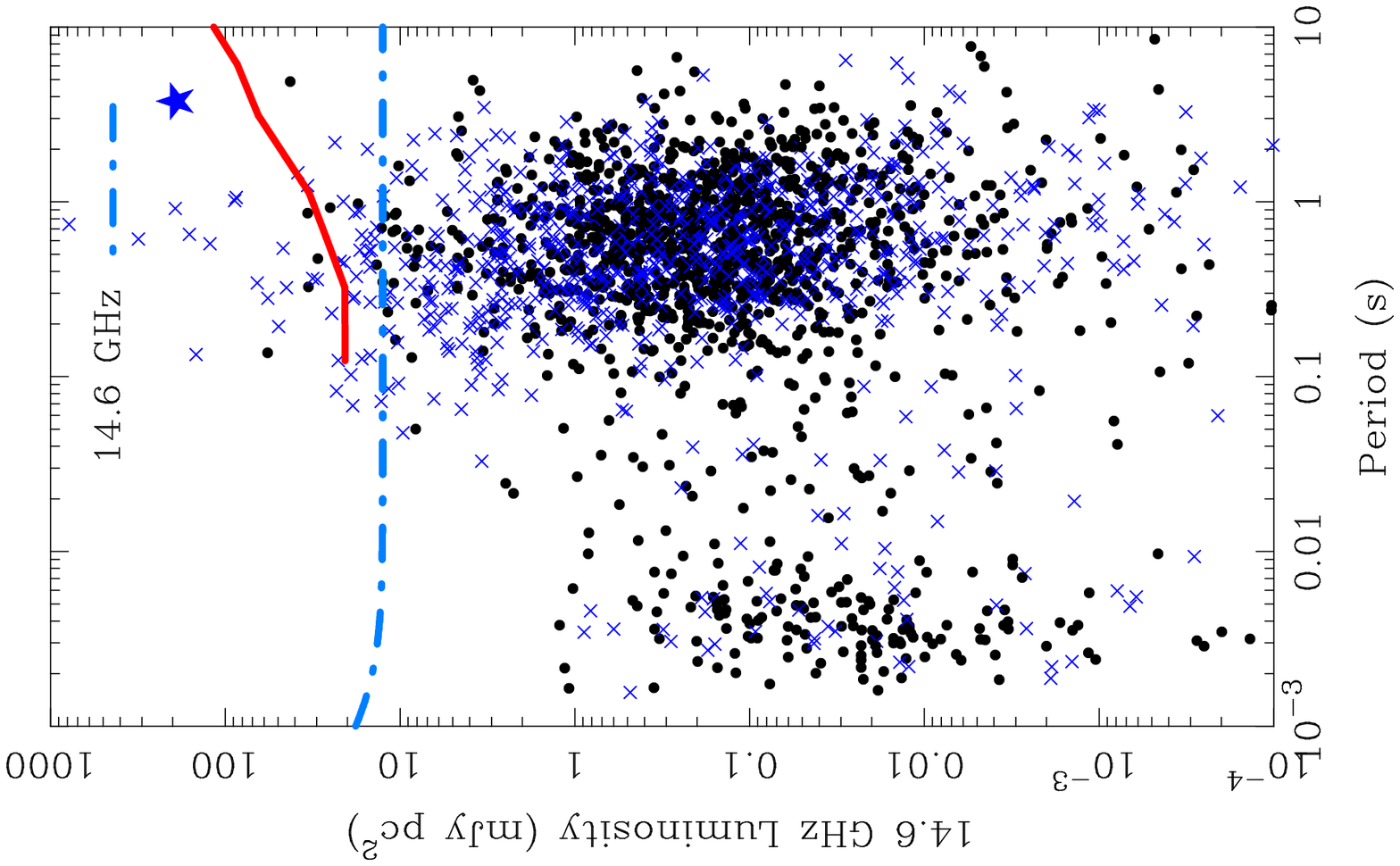}
    \hspace{1.2cm}
    \includegraphics[height=6.5cm,angle=-90]{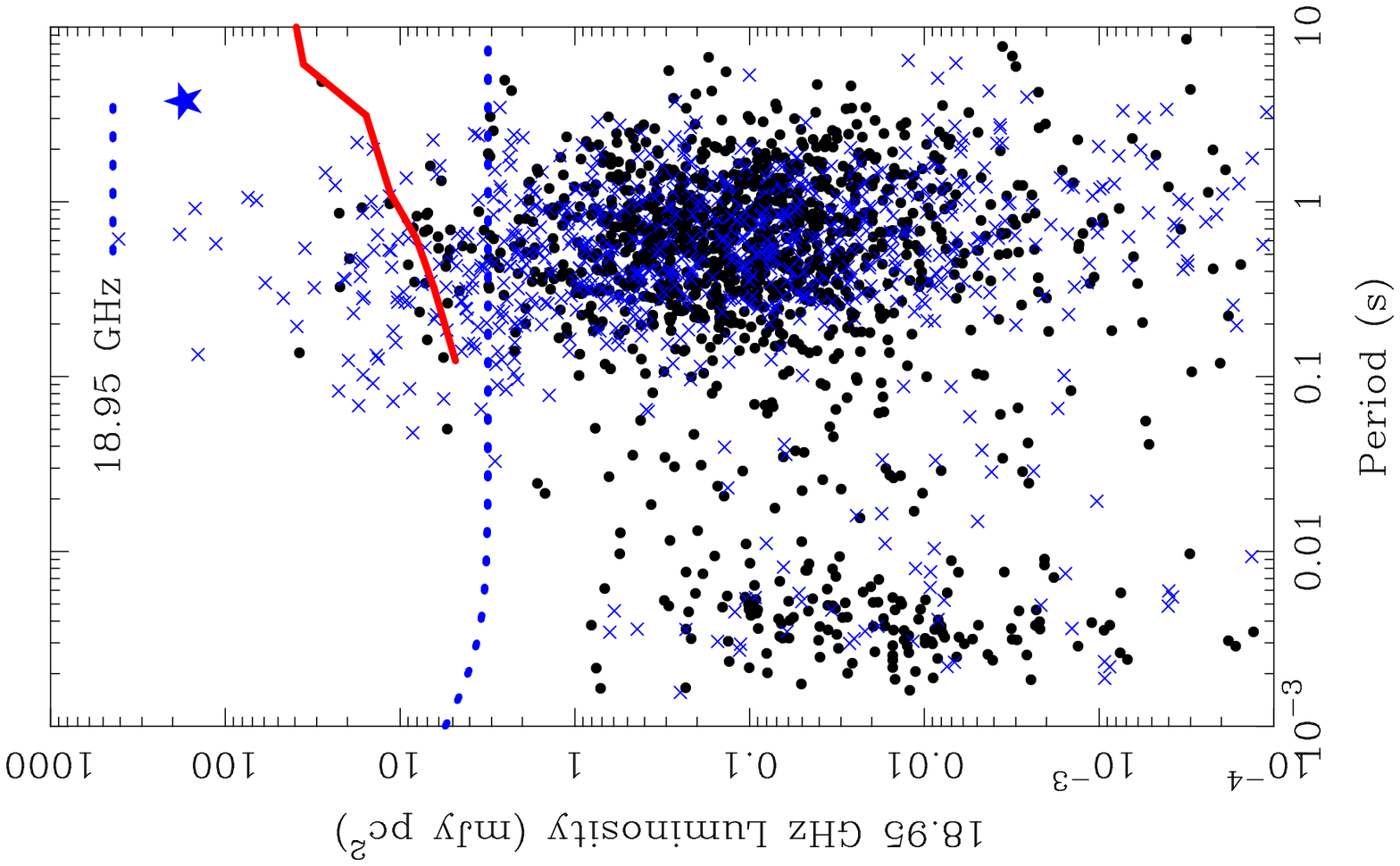}
    \caption{Luminosity versus spin period of the known pulsar population at $14.6\,{\rm GHz}$ (left hand panel) and $18.95\,{\rm GHz}$ (right hand panel). In these diagrams the pulsar luminosity is scaled from either $1.4\,{\rm GHz}$ and $0.4\,{\rm GHz}$ using the known spectral index (pulsars with an index are marked with blue crosses) or a spectra index of $-1.6$. The estimate of the $10\sigma$ sensitivity limit of our GC search observation (for the maximum observing duration) is marked with a line. The red line indicates measured $10\sigma$ sensitivity limits from the injection of simulated pulsar signals which account for red noise effects. See Appendix~\ref{s:appendixB} for more details about this limit.} 
    \label{f:lum_sens_app2}
    \end{center}
\end{figure*}

In this section we outline the recovered fractions of a hypothetical GC pulsar population after scaling the known pulsar population flux densities (and corresponding luminosity) to each individual observing frequency presented in this work. At each reference frequency we have used both the known spectral index and either a simple average spectral index of $-1.6$ or a random selection from a Gaussian distribution of spectral indices with mean $-1.6$ and standard deviation of $0.54$ (to model the observed spread in pulsar spectral indices) as given in \citet{jvsk+18}. The different frequency scaling types are listed in Table~\ref{t:rec_frac}. At $1.4\,{\rm GHz}$ the majority of pulsars have a known flux density and are therefore not frequency scaled. The effects of these alternative analyses changes the recovered fractions by just one to three percentage points at $4.85$ and $8.35\,{\rm GHz}$. At our highest observing frequencies of $14.6$ and $18.95\,{\rm GHz}$, the recovered fractions of a population can increase by just over a factor of two compared to the same analysis with a reference frequency of $1.4\,{\rm GHz}$.
For high frequency pulsar searches we therefore suggest these analyses should be performed at the observing frequency being used (see e.g. \citealt{tde+21,lde+21}), however for most pulsars it is still not known if spectral indices are constant over wide frequency ranges (see e.g. \citealp{kxj+96}). The alternative period luminosity diagrams corresponding to the analyses are presented in Figures~\ref{f:lum_sens_app1} and \ref{f:lum_sens_app2}. These figures also contain luminosity limits that account for measured red noise effects (red lines) above spin periods of $0.1\,{\rm s}$ - see Appendix~\ref{s:appendixB} for details. Because of such effects, the recovered fractions of GC pulsars given here should be treated as approximate upper limits.             

\begin{table*}
  \centering
    \caption{The estimated recovered fractions of a hypothetical GC pulsar population by the searches conducted in this work (and displayed in Figures~\ref{f:lum_sens}, \ref{f:lum_sens_app1} and \ref{f:lum_sens_app2}). $2125$ pulsars with flux density measurements at either $1.4\,{\rm GHz}$ or $0.4\,{\rm GHz}$, known or unknown spectral indices and distance estimates from version 1.62 of the ATNF pulsar catalogue form the sample. Each column can be summarised as follows: ``Ref. frequency'' indicates the reference frequency at which the luminosity is given and plotted; ``PSR scaling index type'' is how the luminosity of pulsars, with both known and unknown spectral indices, are scaled to the reference frequency (see table footnote for further details and we note that no scaling was required at $1.4\,{\rm GHz}$ for pulsars with known $S_{\rm 1400}$); ``Pulse width'' gives the assumed pulse width as a fraction of the pulse period. At 1.4 GHz all luminosity limits are frequency scaled assuming an average pulsar spectral index of $-1.6$ as given in \citet{jvsk+18}. In all cases a single power law function is assumed. Numbers in parentheses account for the reduction in sensitivity caused by red noise effects described in Appendix~\ref{s:appendixB}.}
  \label{t:rec_frac}
  \begin{tabular}{rccccccc}
    \hline  
    \multicolumn{1}{c}{Ref. frequency} & \multicolumn{1}{l}{PSR scaling} &  \multicolumn{1}{l}{Pulse} & $4.85\,{\rm GHz}$	& $8.35\,{\rm GHz}$ & $14.6\,{\rm GHz}$ & $18.95\,{\rm GHz}$ & Associated\\
    & \multicolumn{1}{l}{index type} & \multicolumn{1}{l}{width} &&&&&\multicolumn{1}{l}{figure} \\
    \hline
    \hline
    $1.4\,{\rm GHz}$ & (a)  & $0.05P$     & $11\%$         &  $10\%$   &  $1\%$ & $4\%$ & Fig.~\ref{f:lum_sens} \\  
    $1.4\,{\rm GHz}$ & (a)  & $0.10P$     & $8\%$          &  $7\%$    &  $1\%$ & $3\%$ & $-$\\ 
    $1.4\,{\rm GHz}$ & (b)  & $0.05P$     & $12\%$         &  $10\%$   &  $1\%$ & $4\%$ & $-$ \\  
    $1.4\,{\rm GHz}$ & (b)  & $0.10P$     & $8\%$          &  $7\%$    &  $1\%$ & $3\%$ & $-$\\ 
    \hline
    $4.85\,{\rm GHz}$ & (c)  & $0.05P$   & $12(9)\%$          &  $-$       &  $-$         &  $-$     & Fig.~\ref{f:lum_sens_app1}      \\ 
    $4.85\,{\rm GHz}$ & (c)  & $0.10P$   & $9\%$           &  $-$       &  $-$         &  $-$     & $-$    \\ 
    $4.85\,{\rm GHz}$ & (d)  & $0.05P$   & $13\%$           &  $-$       &  $-$         &  $-$     & $-$    \\
    $4.85\,{\rm GHz}$ & (d)  & $0.10P$   & $10\%$           &  $-$       &  $-$         &  $-$     & $-$    \\
    $8.35\,{\rm GHz}$ & (c)  & $0.05P$   & $-$               &  $11(8)\%$  &  $-$         &  $-$     & Fig.~\ref{f:lum_sens_app1}    \\
    $8.35\,{\rm GHz}$ & (c)  & $0.10P$   & $-$               &  $8\%$   &  $-$         &  $-$     & $-$    \\
    $8.35\,{\rm GHz}$ & (d)  & $0.05P$   & $-$               &  $13\%$   &  $-$         &  $-$     & $-$    \\
    $8.35\,{\rm GHz}$ & (d)  & $0.10P$   & $-$               &  $10\%$   &  $-$         &  $-$     & $-$    \\
    $14.6\,{\rm GHz}$ & (c)  & $0.05P$   & $-$               &  $-$       &  $3(1)\%$     &  $-$     & Fig.~\ref{f:lum_sens_app2}    \\
    $14.6\,{\rm GHz}$ & (c)  & $0.10P$   & $-$               &  $-$       &  $2\%$     &  $-$     & $-$    \\
    $14.6\,{\rm GHz}$ & (d)  & $0.05P$   & $-$               &  $-$       &  $4\%$     &  $-$     & $-$    \\
    $14.6\,{\rm GHz}$ & (d)  & $0.10P$   & $-$               &  $-$       &  $3\%$     &  $-$     & $-$    \\
    $18.95\,{\rm GHz}$ & (c) & $0.05P$   & $-$               &  $-$       &  $-$         &  $6(4)\%$ & Fig.~\ref{f:lum_sens_app2}\\
    $18.95\,{\rm GHz}$ & (c) & $0.10P$   & $-$               &  $-$       &  $-$         &  $5\%$ & $-$ \\
    $18.95\,{\rm GHz}$ & (d) & $0.05P$   & $-$               &  $-$       &  $-$         &  $9\%$ & $-$ \\
    $18.95\,{\rm GHz}$ & (d) & $0.10P$   & $-$               &  $-$       &  $-$         &  $6\%$ & $-$ \\

    \hline
     \end{tabular}
     \\\footnotesize{PSR scaling index types: (a) none and $-1.6$; (b) none and rand.; (c) known and $-1.6$; (d) known and rand.}
\end{table*}

\section{The effects of low fluctuation frequency noise on sensitivity}
\label{s:appendixB}
\begin{figure*}
    \begin{center}
    \includegraphics[height=12.5cm,angle=-90]{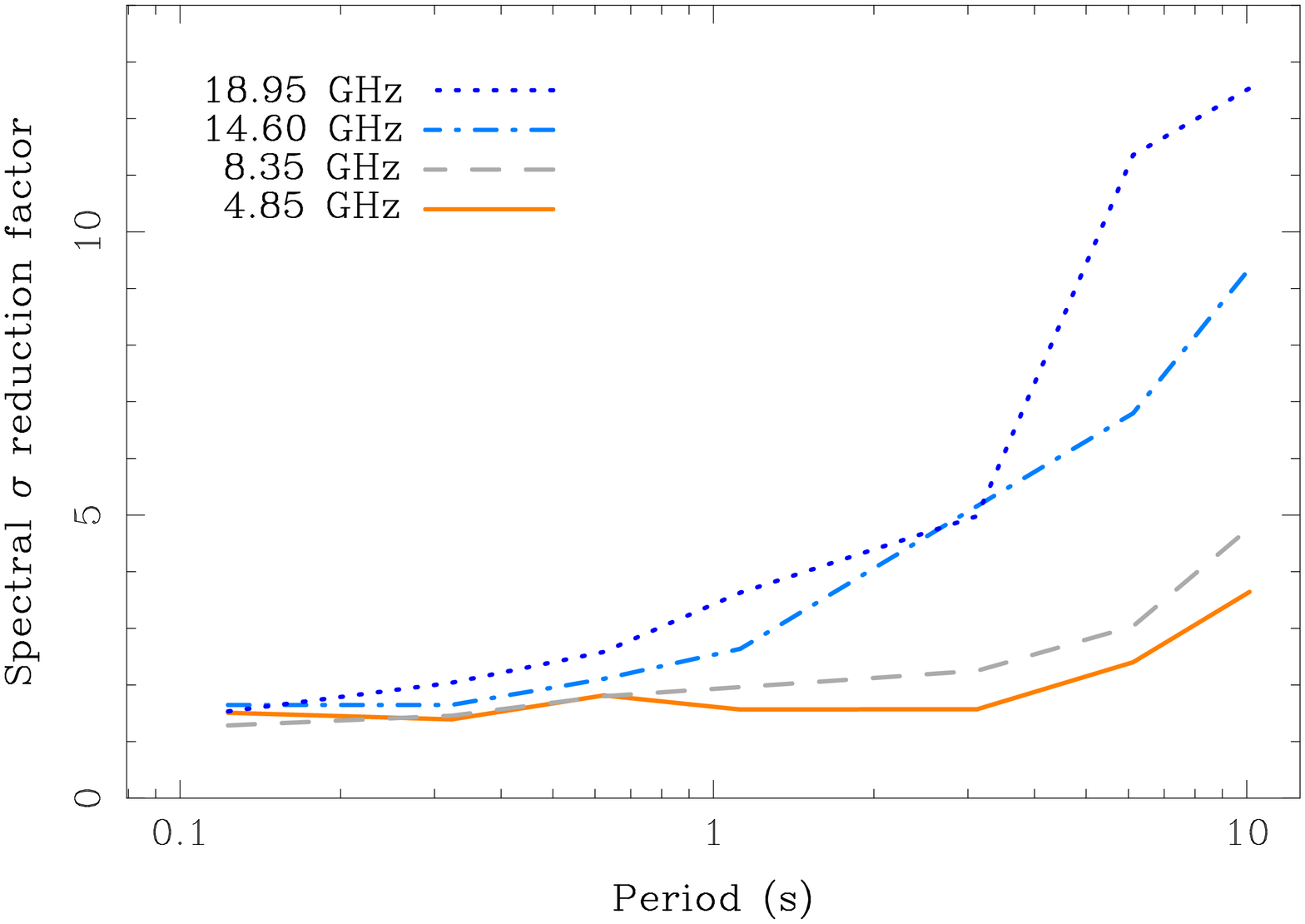}
    \caption{The measured reduction factor of spectral $\sigma$ given in {\sc accelsearch} due to red noise effects as a function of spin period. See Appendix~\ref{s:appendixB} for details of how the reduction factor was measured.} 
    \label{f:red_noise_eff}
    \end{center}
\end{figure*}

It has been shown that a combination of RFI and low frequency (viz. fluctuation frequency) noise variations (red noise) in the $1.4\,{\rm GHz}$ radio observing system of the Arecibo Pulsar ALFA (PALFA) survey, adversely impacts the sensitivity to longer period ($P\gtrsim0.1\,{\rm s}$) pulsars \citep{lbh+15}. At the observing frequencies presented in this work, we detect considerably less RFI than Effelsberg observations at $1.4\,{\rm GHz}$, however RFI is still present and red noise effects might occur due to atmospheric opacity variations and/or receiver/backend fluctuations. To examine these effects we have performed the following tests: 

Simulated pulsar signals have been injected into examples of both the real observational data and simulated data consisting of purely Gaussian white noise. The latter forms our ``baseline'' from which we can measure  sensitivity losses empirically. We have made use of the {\sc presto} routine {\sc makedata} to simulate both noise free pulsar signals and time series of purely Gaussian white noise. Firstly we take an example dedispersed time series (dedispersed to the DM of PSR~J1745$-$2900 of $1778\,{\rm cm}^{-3}\,{\rm pc}$) at a given frequency of either $4.85,\,8.35,\,14.60$ or $18.95\,{\rm GHz}$. The ``DC offset'' of this time series (effectively the value of $T_{\rm sys}$ in un-calibrated machine counts) is then measured in order to simulate a time series with equivalent standard deviation fluctuations (assuming Poisson statistics) after running the procedure of red noise removal with the {\sc presto} routine {\sc rednoise}, as is done in our pipeline (Section~\ref{GCsurvey:sec_data_analysis}).  The noise free pulsar signal of a period $P$ and width $0.05P$ is then injected into the simulated white noise data where the spectral detection $\sigma$ value is measured with {\sc accelsearch}. The same simulated noise free pulsar signal is then injected into the real survey data after which red nose removal is performed and the spectral detection $\sigma$ is also measured with {\sc accelsearch}. We then find the relative reduction factor in spectral $\sigma$ as a function of spin period. Non-integer spin periods of $0.123,\,0.323,\,...\,10.123\,{\rm s}$ have been simulated in order to minimise coincidences with terrestrial RFI signals (which often occur around integer spin periods) that could bias results. The reduction factor in {\sc accelsearch} spectral $\sigma$ is plotted for each frequency in Figure~\ref{f:red_noise_eff}.  Note the apparent worsening of effects with increasing observing frequency. This might be caused by the increased effects of atmospheric opacity variations at higher frequencies. 

The reduction factor in spectral $\sigma$ allows us to scale the $10\sigma$ luminosity survey limits as a function of spin period and is plotted with red lines in Figures~\ref{f:lum_sens_app1} and \ref{f:lum_sens_app2}. The effects on the recovered fraction of a GC pulsar population are also indicated in Table~\ref{t:rec_frac} by the numbers given in parentheses. Recovered fractions are reduced by a few percentage points at all frequencies with proportionately the biggest effects at at $14.6$ and $18.95\,{\rm GHz}$. Further in depth multi-epoch analyses are required to fully account for changing atmospheric conditions.

\bsp	
\label{lastpage}
\end{document}